\documentclass[twocolumn,showpacs,aps,epsfig,nofootinbib]{revtex4}

\usepackage{graphicx}
\usepackage{epstopdf}
\usepackage{latexsym}
\usepackage{amssymb}
\usepackage{color}

\usepackage[center]{subfigure}

\begin{document}

 \newcommand{\bq}{\begin{equation}}
 \newcommand{\eq}{\end{equation}}
 \newcommand{\bqn}{\begin{eqnarray}}
 \newcommand{\eqn}{\end{eqnarray}}
 \newcommand{\nb}{\nonumber}
 \newcommand{\lb}{\label}
\newcommand{\PRL}{Phys. Rev. Lett.}
\newcommand{\PL}{Phys. Lett.}
\newcommand{\PR}{Phys. Rev.}
\newcommand{\CQG}{Class. Quantum Grav.}

\title{Universal horizons and black holes in gravitational theories with broken Lorentz symmetry}

\author{Kai Lin $^{a, b}$}

\author{Elcio Abdalla $^{b}$}

\author{Rong-Gen Cai $^{c, d}$}

\author{Anzhong Wang $^{a, e}$\footnote{The corresponding author\\ E-mail: Anzhong$\_$Wang@baylor.edu}}

\affiliation{$^{a}$  Institute  for Advanced Physics $\&$ Mathematics, Zhejiang University of Technology, Hangzhou 310032,  China\\
$^{b}$  Instituto de F\'isica, Universidade de S\~ao Paulo, CP 66318, 05315-970, S\~ao Paulo, Brazil \\
$^{c}$  State Key Laboratory of Theoretical Physics,
Institute of Theoretical Physics, Chinese Academy of Sciences, Beijing 100190, China\\
$^{d}$ King Abdulaziz University, Jeddah 21589, Saudi Arabia\\
 $^{e}$ GCAP-CASPER, Physics Department, Baylor University, Waco, TX 76798-7316, USA }

\date{\today}

\begin{abstract}

In this paper, we first show that the definition of the universal horizons studied  recently in the  khronometric theory of gravity can be
straightforwardly generalized to other theories  that violate  the Lorentz symmetry, by simply considering  the khronon  as a probe field
and playing  the same role as a Killing vector field. As an application, we study static charged ($D+1$)-dimensional spacetimes in
the framework of the healthy (non-projectable)  Horava-Lifshitz (HL)  gravity in the infrared limit, and find various solutions.  Some of
them represent Lifshitz space-times with hyperscaling violations, and some have black hole structures. In the latter universal horizons
always exist inside the Killing horizons. The surface gravity on  them   can be either larger or smaller than the surface gravity on the
Killing horizons, depending  on the space-times  considered.  Although such black holes are found only in the infrared, we argue that
black holes  with universal horizons also exist in the full theory of the HL gravity. A simple example is the Schwarzschild solution written
in  the  Painleve-Gullstrand  coordinates, which is also a solution of the  full theory of the  HL  gravity and has a universal horizon located
inside the Schwarzschild Killing horizon.

\end{abstract}


\maketitle

\section{ Introduction  }
\renewcommand{\theequation}{1.\arabic{equation}} \setcounter{equation}{0}

Lorentz symmetry has been the cornerstone of modern physics,   and verified to  such a high accuracy  that any modification
of it must face one of the most severe experimental constraints existing today in physics \cite{Liberati13}, although it is arguable that such
constraints in the matter sector are much stronger than those in the gravitational sector  \cite{LZbreaking}.
Nevertheless, if space and time emerge from some discrete substratum, as what we currently understand, this symmetry must be an
accidental one at low energies.

Following this line of thinking, various gravitational theories that violate Lorentz symmetry have been proposed,
ranging from ghost condensation \cite{GC},  Einstein-aether theory \cite{EA},  and more recently, to  Horava-Lifshitz (HL)  gravity \cite{Horava}.
While the  ghost condensation and Einstein-aether theory are all considered as the low energy effective theories of gravity,
the HL gravity is attempted to be ultraviolet (UV) complete, and by construction  is power-counting renormalizable \cite{reviews}.
It is consistent with all the solar system tests carried out so far \cite{EA,KMWZ} and binary pulsar observations \cite{Yagi}, and exhibits various remarkable features when applied to cosmology
\cite{InflationA,InflationB,InflationC,InflationD}.

However, when applying the HL theory to astrophysics, it seems to indicate that black holes are only low energy phenomena \cite{KKCP,GLLSW}.
This is because, in order to have the theory power-counting renormalizable, high-order spatial operators up to at least sixth-order must be included \cite{Horava}.
Hence, the dispersion relation becomes nonlinear, and generically takes the form \cite{reviews},
\bq
\lb{1.1}
E^2 = c_{p}^2 p^2\left(1 + \alpha_1 \left(\frac{p}{M_{*}}\right)^2 +  \alpha_2  \left(\frac{p}{M_{*}}\right)^4\right),
\eq
where $E$ and $p$ are the energy and momentum of the particle considered, and $c_p, \; \alpha_i$ are coefficients, depending on the particular
specie of the particle, while $M_{*}$ denotes the suppression energy  scale of the higher-dimensional operators. Then,   one can see that
both phase and group velocities of the particles are unbounded  with the increase of energy.  This makes the causal
structure of the spacetimes quite different from that given in general relativity (GR),  where the light cone of a given point  plays a fundamental
role in determining its causal structure relative to other events.   In the case described by Eq.(\ref{1.1}),   the causal
structure is very much similar to the Newtonian case \cite{GLLSW}.  This suggests that black holes may not exist at all in the
HL theory, as any ray initially trapped inside a horizon can penetrate it and propagate to infinity, as long as the ray has sufficiently large velocity. On
the other hand, in the infrared (IR) the high-order terms
are negligible, and the first term in Eq.(\ref{1.1}) becomes dominant, so one may still define
black holes, following what was done  in GR \footnote{Even in this limit, there are subtles in defining black holes. For example, in the Einstein-aether theory spin-1 and spin-0 gravitons
exist \cite{EA}. To avoid  the Cherenkov effects \cite{EMSa}, these particles must propagate with speeds no less than that of light. Clearly, they can penetrate the event horizons to escape to infinities,
even these particles  are initially trapped inside the Killing horizons. }.

Surprisingly, in contrast to the above physical intuition, recently
it was shown that there still exist absolute causal boundaries, the
so-called {\em universal horizons}, and particles even with
infinitely large velocities     would just move around on these
boundaries and  cannot escape to infinity \cite{BS11}. This has
immediately attracted lot of attentions
\cite{UHsA,UHsB,BBM,CLMV,SVV}. In particular,  it was shown that the
universal horizon radiates like a blackbody at a fixed temperature,
and obeys  the first law of black hole mechanics  \cite{BBM}. The
main idea is as follows: In a given space-time, a timelike foliation
parametrized by $\phi\left(x^{\mu}\right) = $ Constant might exist
globally. Since the surfaces are timelike, we must have $u^{\lambda}
u_{\lambda} = - 1$, where $u_{\mu}$ is the normal unit vector of the
surface, defined as
 \bq
\lb{1.2}
u_{\mu} = \frac{\phi_{,\mu}}{\sqrt{X}},
\eq
with $ \phi_{,\mu} \equiv \partial \phi/\partial x^{\mu}, \; X \equiv -g^{\alpha\beta}\partial_{\alpha} \phi \partial_{\beta} \phi$. The signatures of the metric are $(-1, 1, ..., 1)$.
Among these surfaces,  there may exist a   surface at which $\phi$ diverges, while physically nothing
singular happens there, including  the metric and the space-time. Given that $\phi$ defines an absolute time, any object crossing this surface from the interior would necessarily also move back in
absolute time, which is something forbidden by the definition of the causality in the theory. Thus, even particles with superluminal velocities cannot penetrate this surface, once they are
trapped inside it. For more details, we refer readers to \cite{BS11}.

In this paper, our purposes are twofold: First, we shall generalize
the above definition of the universal horizons to any gravitational
theory that may or may not violate the Lorentz symmetry, although
such a generalization might be useful only for theories that violate
the Lorentz symmetry.  In \cite{BS11},   $\phi\left(x^{\mu}\right) $
was referred to as the ``khronon" field, and considered as
describing one degree of freedom of  the gravitational field, a
spin-0 graviton.  To generalize the definition of the universal
horizons to other theories, in this paper we shall promote it to the
same role as played by a Killing vector of a given space-time, so
its existence does not affect the background, but defines the
properties  of a given space-time. By this way, such a field is no
longer part of the underlaid gravitational theory and it may or may
not exist in a given space-time, depending on the properties of the
space-time considered. Second, we shall find  static charged
solutions of the healthy extensions of the HL gravity \cite{BPS}  to show that the universal horizons
exist in some of these solutions. Such horizons exist not only in
the IR limit of the HL gravity, as has been considered  so far in
\cite{BS11,UHsA,UHsB,BBM,CLMV,SVV}, but also in the full HL gravity,
that is, when high-order operators are not negligible.

The rest of the paper is organized as follows:  In Sec. II, we generalize the definition of the universal horizons first discovered in \cite{BS11} in the khronometric theory, which is
equivalent to  the Einstein-aether theory with the hypersurface-orthogonal condition \cite{Jacobson10,Wang13}, to any theory by considering the khronon field as a probe field, quite similar to
a Killing vector field existing in a given space-time, so that whether   a  khronon field exists or not depends totally on the  properties of a given space-time. In Sec. III, we consider
charged static spacetimes in the framework of the non-projectable HL gravity in the IR limit, and present various classes of solutions. In Sec. IV, we study the existence of universal horizons in some of the
solutions presented in Sec. III, and find that universal horizons indeed exist. The paper is ended with Sec. V, in which we derive our main conclusions and present some discussing remarks.
An appendix is also included, in which we briefly review the non-projectable HL gravity in ($D+1$)-dimensional spacetimes when coupled with an electromagnetic field.

It should be noted that electromagnetic static spacetimes in other versions of the HL theory were studied in \cite{EMa}, while a new mechanism for generation of primordial magnetic seed field in the early
universe without the local U(1) symmetry was considered in \cite{EMb}. However, so far  no studies of the existence of the universal horizons have been carried out in these models.

\section{ Causal Structure of Gravitational Theories with Broken  Lorentz Symmetry and Universal Horizons   }

\renewcommand{\theequation}{2.\arabic{equation}} \setcounter{equation}{0}

\begin{figure}[tbp]
\centering
\includegraphics[width=8cm]{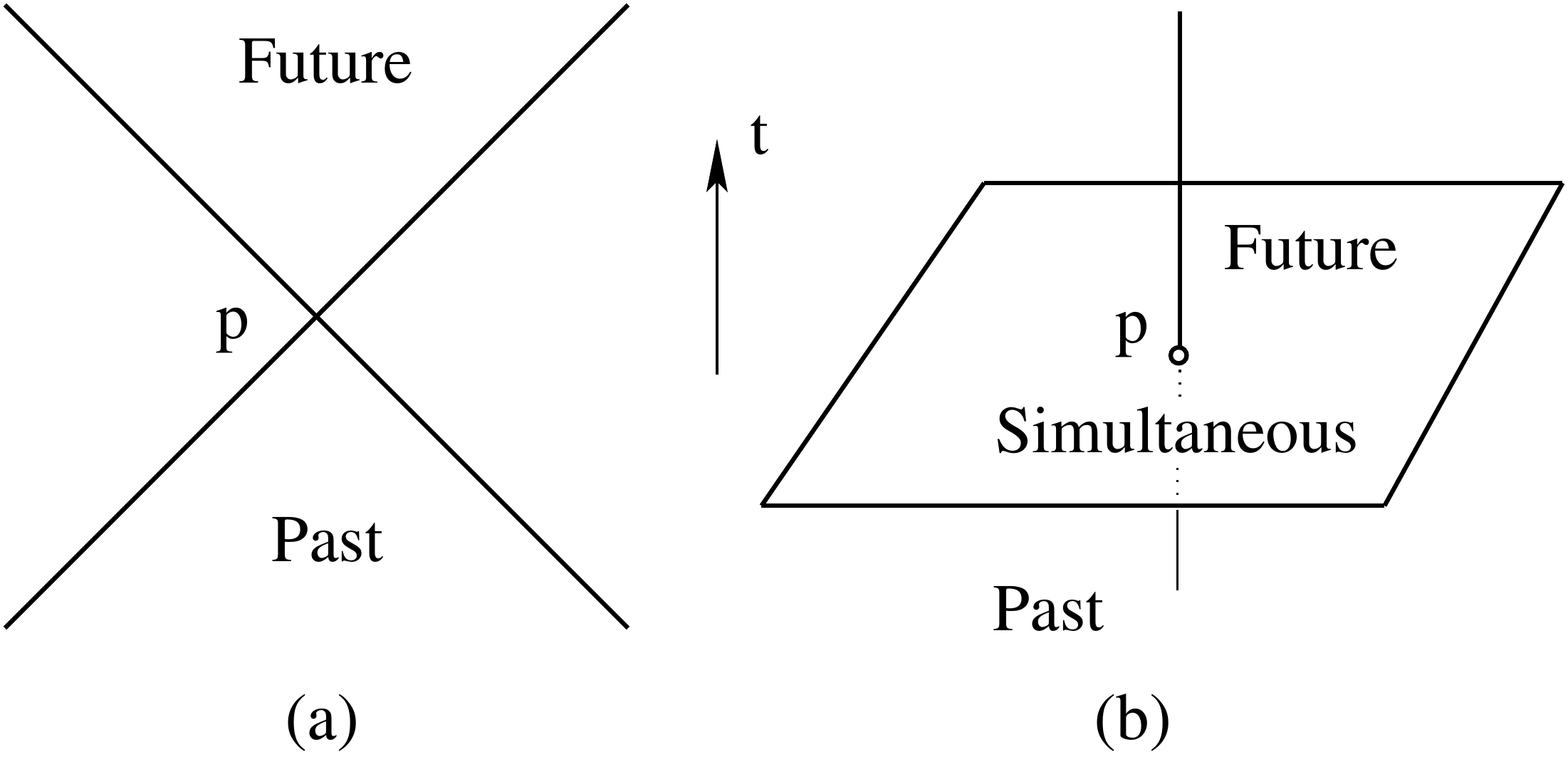}
\caption{ (a) The light cone of the event $p$ in GR. (b) The causal structure of the event $p$  in the HL gravity (See also \cite{GLLSW}).}
\label{fig0}
\end{figure}

As shown in the last section, once the Lorentz symmetry is broken, the speed of particles can become  superluminal, and even instantaneous propagation exists.
Then, the causal structure will be quite different from that in theories with Lorentz symmetry, in which light-cones play a central role. It should be noted that the violation
of the Lorentz symmetry does not mean the violation of the causality. In fact, in such theory the causality still exists, but different from that given, for example, in GR. In particular,
now  the causal structure of  a given point $p$ is uniquely determined by the time difference, $\Delta{t} \equiv t_{p} - t_{q}$, between the two events $p$ and $q$.
 If $\Delta{t} > 0$, the event $q$ is to the past of $p$; if $\Delta{t} < 0$, it  is to the future; and if $\Delta{t} = 0$, the two events are
simultaneous [cf. Fig.\ref{fig0}].

As a result,  all the  definitions of black holes in terms of event
horizons \cite{HE73,Tip77,Hay94,Wang} become invalid, as  a particle
initially trapped inside such a horizon now can penetrate  it and
propagate to infinity, as long as its velocity is sufficiently
large.  To provide a proper definition of black holes, anisotropic
conformal boundaries \cite{HMT2} and  kinematics of particles
\cite{KM} have been studied    in the framework of the HL gravity.
In particular, defining  a horizon  as the infinitely redshifted
2-dimensional (closed) surface of massless test particles
\cite{KKb}, it was found that for test particles with sufficiently
high  energy, the radius of the horizon can be made as small as
desired, although the singularities can be seen in principle only by
observers with  infinitely high energy \cite{GLLSW}. This is
expected, as such horizons are similar to the event horizon defined
in GR \cite{HE73}.

Remarkably, studying the behavior of a khronon field in the fixed Schwarzschild black hole background,
\bq
\lb{eq1.1}
ds^2 = - \left(1- \frac{r_s}{r}\right)dv^2 + 2dvdr + r^2d\Omega^2,
\eq
where $r_s \equiv 2M$,  Blas and Sibiryakov showed that a universal horizon exists
inside the Schwarzschild radius $r_s$ \cite{BS11} \footnote{In \cite{BS11} the authors assumed that the coupling constants are very much smaller than
unity, so that the backreaction of the khronon field to the space-time is negligible, and in this limit the Schwarzschild solution is also  a solution of the khronometric theory. }. This surface, in contrast to the event horizon, now is spacelike, and on which the time-translation  Killing
vector $\zeta^{\mu} =   \delta^{\mu}_{v}$ becomes orthogonal to    $u^{\mu}$,
\bq
\lb{eq1.2}
u_{\mu} \zeta^{\mu} = 0.
\eq
Since $u_{\mu}$ is well-defined in the whole space-time, and remains timelike from the asymptotical infinity ($r = \infty$) all the way down to the space-time singularity ($r = 0$),
Eq.(\ref{eq1.2}) is possible only inside the Killing horizon ($\zeta^{\lambda}\zeta_{\lambda} = 0$), as only there  $\zeta^{\mu}$ becomes spacelike and can be
possibly orthogonal to $u_{\mu}$.

\begin{figure}[tbp]
\centering
\includegraphics[width=8cm]{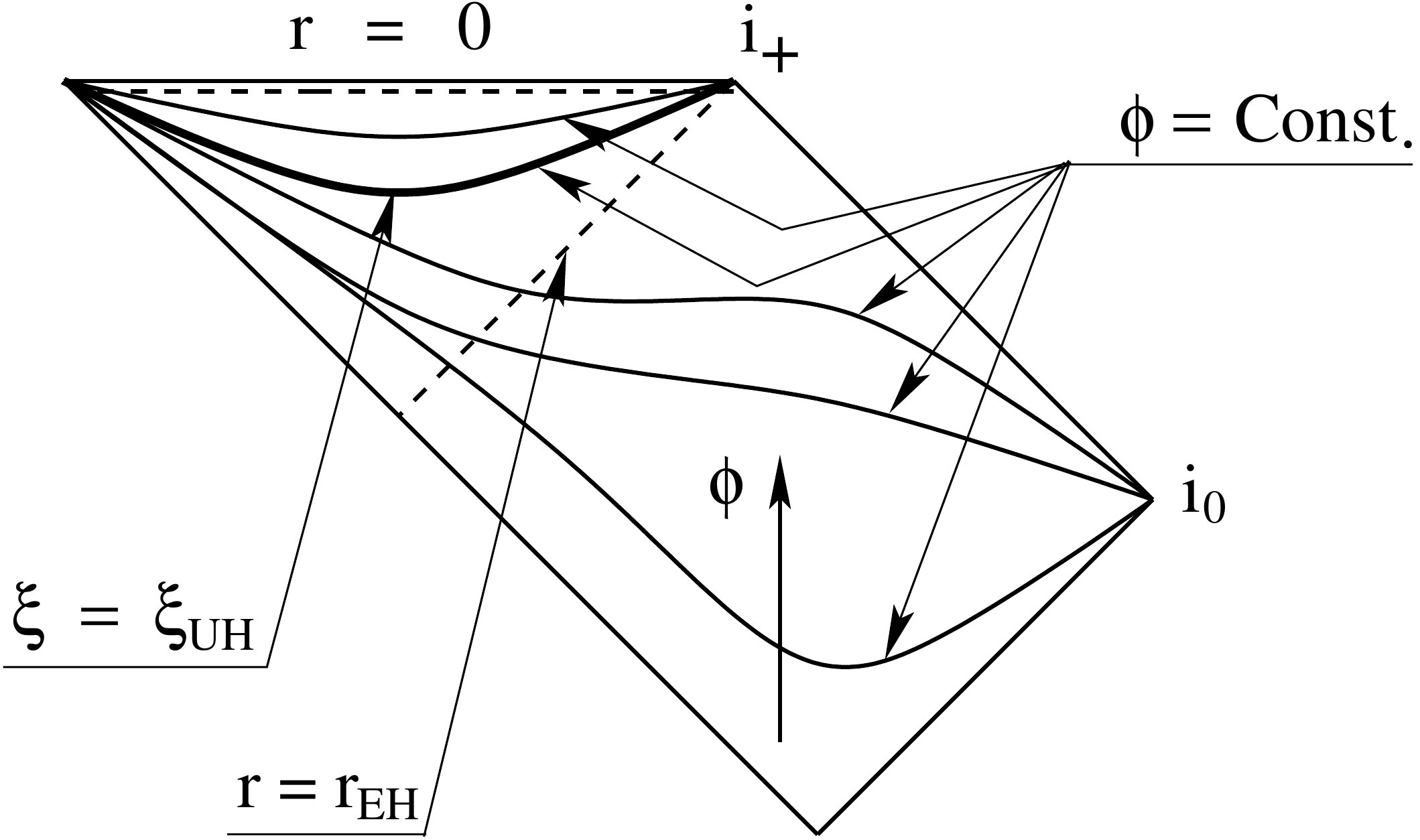}
\caption{ The foliation of the timelike hypersurfaces $\phi = $
Constant, and the location of the universal horizon $\xi = \xi_{UH}$
on the  ($v, r$)-plane. The  khronon defines globally an absolute
time, and the trajectory of a particle is always along the
increasing direction of $\phi$. Thus, once it cross the horizon, the
particle move toward the singularity $r = 0$ and reaches it within a
finite proper time.} \label{fig00}
\end{figure}

The  khronon defines globally an absolute time, and the trajectory of a particle must be always along the increasing direction of $\phi$.  Thus,
once the particle across the universal horizon, the only destination is to move towards the space-time singularity, and arrive at it within a finite (proper) time,
as shown in Fig.\ref{fig00}. From this same figure, one can also see that the normal vector $u_{\mu}$ is pointing outwards  for some $\phi$ at the event horizon
$r = r_{EH}$. That is, for a particle with a sufficient large velocity (larger than that of light), it can escape from the interior  of the event horizon to asymptotically-flat infinities.
In particular, near the universal horizon the  khronon  behaves like \cite{BS11},
 \bq
 \lb{eq1.3}
 \phi \simeq  v + \frac{\log(\xi_{UH} - \xi)}{\xi_{UH}^2u_{\tau}'\sqrt{\xi_{UH}^2 -1}},
 \eq
 where $\xi \equiv r_s/r$, and $\xi = \xi_{UH}$ (or $\phi = \infty)$ is the location of the universal horizon. $u_{\tau}$ is the $\tau$-component of the  khronon field, and $u_{\tau}' \equiv du_{\tau}/d\xi$.
 Here
$\tau$ is the Schwarzschild timelike coordinate, defined as  \cite{BS11}
 \bq
 \lb{eq1.4}
 \tau = v - \left[r + r_s\ln\left(\frac{r}{r_s} - 1\right)\right].
 \eq
The hypersurfaces of $\phi = $ Constant are illustrated in Fig.\ref{fig00}, from which we can see that these curves are all accumulated to the one $\phi = \infty\; (\xi = \xi_{UH})$,
which is  the location of the universal horizon
and only particles with infinitely large velocities can move around on this surface.
A particle inside  this surface cannot get out of it, no matter how large its velocity would be.

It should be noted that the singularity of the  khronon on the universal horizon is not physical, and can be removed by the gauge transformation,
\bq
\lb{1.5}
\tilde{\phi}  =  {\cal{F}}(\phi),
\eq
allowed by the symmetry of the khronon field, where  ${\cal{F}}(\phi)$ is an arbitrary monotonic function of $\phi$. In this sense, the khronon is quite different from a usual  scalar field.

To generalize the above definition of the universal horizon to other gravitational  theories, one can see that two important ingredients are essential: the existence of the khronon field
$\phi\left(x^{\mu}\right)$, and the existence of the  asymptotically timelike Killing vector $\zeta^{\mu}$. For a given space-time, the latter can be obtained by solving the Killing equation,
\bq
\lb{eq1.6}
D_{\nu}\zeta_{\mu} + D_{\mu}\zeta_{\nu} = 0,
\eq
where $D_{\mu}$  denotes the covariant derivative with respect to the ($D+1$)-dimensional metric $g_{\mu\nu}$.

To find the equation for the  khronon, we start with its general action \cite{EA},
\bqn
\lb{eq1.7}
S_{\phi} &=&  \int d^{D+1}x \sqrt{|g|}\Big[c_1\left(D_{\mu}u_{\nu}\right)^2 + c_2 \left(D_{\mu}u^{\mu}\right)^2\nb\\
&& ~~  + c_3   \left(D^{\mu}u^{\nu}\right)\left( D_{\nu}u_{\mu}\right)   - c_4 a^{\mu}a_{\mu} \Big],
\eqn
where $a_{\mu} \equiv u^{\alpha}D_{\alpha}u_{\mu}$, and $c_i$'s are arbitrary constants.
It should be noted that the above action is the most general one in the sense that the
resulting  differential equations in terms of   $u_{\mu}$ are  second-order \cite{EA}. However, when $u_{\mu}$ is hypersurface-orthogonal, that is, when $u_{\mu}$
satisfies the relation,
\bq
\lb{eq1.7a}
u_{[\nu}D_{\alpha}u_{\beta]} = 0,
\eq
only three of them are independent \footnote{Note that with the condition (\ref{eq1.7a}),  $u_{\mu}$ can be always written in the form (\ref{1.2}) \cite{Wald}.},
as in this case we have the identity  \cite{EA},
\bq
\lb{eq1.7b}
\Delta{\cal{L}}_{\phi} \equiv  a^{\mu}a_{\mu} - \big(D_{\alpha}u_{\beta}\big)\big(D^{\alpha}u^{\beta}\big) +   \big(D_{\alpha}u_{\beta}\big)\big(D^{\beta}u^{\alpha}\big) = 0.
\eq
Then, one can always add the term,
\bq
\lb{1.11a}
\Delta{S}_{\phi} = \alpha \int{\sqrt{|g|} \; d^{D+1}x \Delta{\cal{L}}_{\phi}},
\eq
into $S_{\phi}$, where $\alpha$ is an arbitrary constant. This is effectively to shift the coupling constants $c_i$ to ${c}_i'$, where
\bq
\lb{1.11b}
{c}_{1}' = c_1 + \alpha,\; {c}_{2}' = c_2,\; {c}_{3}' = c_3  -  \alpha,\;
{c}_{4}' = c_4  - \alpha.
\eq
Thus, by properly choosing $\alpha$, one can always set one of $c_{1, 3, 4}$ to zero. However, in the following we shall leave this possibility open.

Hence, the variation of  $S_{\phi}$ with respect to $\phi$ yields
the khronon equation,
\bqn
\lb{eq1.8}
D_{\mu} {\cal{A}}^{\mu}  = 0,
\eqn
where \cite{Wang13} \footnote{Notice the difference between the signatures of the metric chosen in this paper and the ones in \cite{Wang13}.},
\bqn
\lb{eq1.9}
{\cal{A}}^{\mu} &\equiv& \frac{\left(\delta^{\mu}_{\nu}  + u^{\mu}u_{\nu}\right)}{\sqrt{X}}\AE^{\nu},\nb\\
\AE^{\nu} &\equiv& D_{\gamma} J^{\gamma\nu} + c_4 a_{\gamma} D^{\nu}u^{\gamma},\nb\\
J^{\alpha}_{\;\;\;\mu} &\equiv&  \big(c_1g^{\alpha\beta}g_{\mu\nu} + c_2 \delta^{\alpha}_{\mu}\delta^{\beta}_{\nu}
+  c_3 \delta^{\alpha}_{\nu}\delta^{\beta}_{\mu}\nb\\
&&  ~~~ - c_4 u^{\alpha}u^{\beta} g_{\mu\nu}\big)D_{\beta}u^{\nu}.
\eqn
Eq.(\ref{eq1.8}) is a second-order differential equation for $u_{\mu}$, and to uniquely determine it, two boundary conditions are needed.
These two conditions can be chosen as follows \cite{BS11}:
(i) One of them is to require it to be
aligned asymptotically with the timelike Killing vector,
\bq
\lb{eq1.10}
u^{\mu} \propto \zeta^{\mu}.
\eq
(ii) The second condition can be that the khronon has a regular future sound horizon, which
  is a null surface of the effective metric \cite{EJ},
\bq
\lb{eq1.10b}
g^{(\phi)}_{\mu\nu} = g_{\mu\nu} - \left(c_{\phi}^2 -1\right)u_{\mu}u_{\nu},
\eq
where $c_{\phi}$ denotes the speed of the khronon \footnote{To avoid the Cherenkov effects,
 in the khronon theory, or more general, the Einstein-aether theory, $c_{\phi}$ is required to be no less than  the speed of light $c_{\phi} > c$ \cite{EA,EMSa}. However, since in the current
 case, the khronon is treated as a probe field, such requirement is not needed.}.

With the above definition of the universal horizon, several comments now are in order.  First, the above definition does not refer to any particular theory of gravity. Therefore, it is applicable to
any theory that violates the Lorentz symmetry, including the Einstein-aether theory \cite{EA} and the HL gravity \cite{Horava}. But, there is a fundamental difference between the khronon  introduced
 in this paper   and the one (a particular aether field with hypersurface-orthognal condition) considered  in \cite{BS11,UHsA,UHsB,BBM}. In this paper, the khronon plays the
 same role as a Killing vector field $\zeta^{\mu}$, both of them describe only some  properties of a given space-time and have no effects  on the given space-time. But, in \cite{BS11,UHsA,UHsB,BBM} the khronon
 was considered as a part of the gravitational field, although in some cases their effects on the gravitational fields were assumed to be negligible.
 Second,   in the literature it has been often considered that
 the Einstein-aether theory with the hypersurface-orthogonal condition   (\ref{eq1.7a}) is equivalent to the   non-projectable HL gravity in the low energy limit. This is incorrect, as the two theory have different
 gauge symmetries, and they are equivalent only with a particular choice of the gauge, the $T$-gauge, in which the aether $u_{\mu}$ is aligned with the time coordinate $t$,  that is, choosing $ t = \phi$,
 as shown explicitly in \cite{Jacobson10,Wang13}. In the following, we shall refer to these coordinates as the $T$-coordinates. In   particular, the  Einstein-aether theory is  gauge-invariant under the general coordinate transformations,
 \bq
 \lb{eq1.10d}
 t = \xi^{0}\left(t', {x'}^k\right),\;\; \; x^i = \xi^{i}\left(t', {x'}^k\right),
 \eq
where $\xi^{\mu}$ are arbitrary functions of their indicated arguments. While the HL gravity is   gauge-invariant only under the foliation-preserving
diffeomorphism,
\bq
 \lb{eq1.10e}
 t = f\left(t' \right),\;\; \; x^i = \xi^{i}\left(t', {x'}^k\right).
 \eq
A typical example is the  Schwarzschild space-time  written in the Painleve-Gullstrand coordinates \cite{PG},
\bq
\lb{eq1.11}
ds^2 = - dt^2 + \left(dr + \sqrt{\frac{r_s}{r}} dt\right)^2   + r^2d\Omega^2.
\eq
This solution  is also a solution of the HL gravity (not only in the IR but also in the UV as now $R_{ij} = 0$ and all high-order operators of $R_{ij}$ vanish) \cite{GLLSW}, but the one given by Eq.(\ref{eq1.1}) is not. This is because  theses two solutions  are connected by the coordinate transformations,
\bq
\lb{eq1.12}
dt = dv - \frac{dr}{1 + \sqrt{\frac{r_s}{r}}},
\eq
which are forbidden by the gauge transformations of Eq.(\ref{eq1.10e}), although they are allowed by the ones of Eq.(\ref{eq1.10d}). Therefore, in the Einstein-aether theory
these two solutions describe the same space-time, but in the HL gravity they do not, as they are not connected by any coordinate transformations allowed by its gauge symmetry
(\ref{eq1.10e}). More examples of this kind can be found in \cite{CW}.

Therefore, the Einstein-aether theory with the hypersurface-orthogonal condition is equivalent to the non-projectable HL gravity in the low energy limit only in the $T$-coordinates, in which the timelike foliations
of the space-time fixed  in the HL gravity coincide with the spacelike hypersurfaces $\phi(x^{\mu})= $ Constant \footnote{Therefore, to find solutions  of the HL gravity, one may start with the
Einstein-aether theory with the hypersurface-orthogonal condition in a proper coordinate system. Once such solutions are found, one can transform these solutions to the T-coordinates, whereby the solutions of
the HL gravity can be read off directly. In particular, all the spherically symmetric solutions of the  Einstein-aether theory are hypersurface-orthogonal \cite{EJ,EA,UHsA}, and when they are written in the T-coordinates,
they are also the solutions of the HL gravity.}.

The equivalence shown in \cite{Jacobson10} is actually the equivalence between the Einstein-aether theory with the hypersurface-orthogonal condition   (\ref{eq1.7a}) and the khronometric theory
\cite{BPS}, as both of them are gauge-invariant under the general covariant coordinate transformations (\ref{eq1.10d}) and have the same degree of freedom. For more details, we refer readers to \cite{Jacobson10,Wang13,BS11}.

As an application of the  universal horizons defined above,  in the next section we shall  find static charged solutions in the framework of  the HL gravity without the
projectability condition in the low energy limit \cite{BPS}. In Sec. V, we   study their local and global properties, by paying particular attention to the existence of the universal horizons.

 \section{Static charged Lifshitz-type Solutions in non-projectable $(D+1)$-dimensional HL Gravity}
\renewcommand{\theequation}{3.\arabic{equation}} \setcounter{equation}{0}

The  non-projectable  HL gravity in (D+1)-dimensional space-time is briefly reviewed in Appendix A, in which all the field equations are derived,
including the generalized Maxwell equations.

In this paper, we shall study static  spacetimes  described by,
 \bqn
\lb{3.2}
&& N  =  r^z f(r),\;\;\; N^i = 0,
 \nb\\
&& g_{ab}dx^adx^b  =  \frac{g^2(r)}{r^2}dr^2 +  r^2\delta_{ij}dx^idx^j,\nb\\
&& A_0 = A_0(r),\;\;\;A_r=A_1(r),\;\;\; A_{i} = 0,
 \eqn
in the coordinates ($t, r, x^k$), where $D = d +1,\; (i, j, k = 1, 2, ..., d)$,  and $z$ is a constant.  The $D-$dimensional Ricci scalar
curvature is given by
 \bqn
\lb{Ricci} R=2d  \frac{rg'(r)}{g^3(r)}-\frac{d(d+1)}{g^2(r)}.
 \eqn

In the IR limit, all the operators higher than order 2 can be safely ignored, as mentioned above. Then, from the Maxwell equations (\ref{MaxwellA}) and (\ref{MaxwellB}), we obtain
 \bqn
\lb{MaxwellEq}
&& A_1 \alpha_0'=0, \\
\lb{MaxwellEqb}
&& \frac{A_0''}{A_0'}-\frac{g'}{g}-\frac{f'}{f}+\frac{d+1-z}{r}=0.
 \eqn
Thus,   $\alpha_0=\alpha_0(A^1A_1)$ must be a constant. Then, from Eqs.(\ref{actionMatter}) and (\ref{potentialMatter}) we can see that now $\alpha_0$
acts like a cosmological constant, and can be absorbed to $\Lambda$. 
Therefore, without loss of the generality, we shall set $\alpha_0 = 0$ in the rest of the paper.
On the other hand, from Eq.(\ref{MaxwellEqb}) we find that,
 \bqn
\lb{A0}
A'_0=\frac{q}{r^{d+1-z}}f(r)g(r),
 \eqn
where $q$ is an integration constant.

Substituting the above ADM variables and $A_0$ into the rest of field equations, we  find that the momentum
constraint vanish directly, but the Hamiltonian constraint and $(r, r)$-
and $(i,i)$-components  of the dynamical equations are non-trivial.
It can be shown that the $(i, i)$-component  of the dynamical equations can be
derived from the Hamiltonian constraint and the $(r, r)$-component. Therefore, in the current case there are only two independent field equations, after integrating out the
electromagnetic field equations (\ref{MaxwellEq}) and (\ref{MaxwellEqb}), which are sufficient to determine the two unknown functions $f(r)$ and $g(r)$.
The Hamiltonian
constraint and the $(r, r)$-component  of the dynamical equations are given, respectively,  by
 \bqn
 \lb{feqA}
&&-2r^2\beta\frac{f''}{f}+\left(2r^2\beta\frac{g'}{g}-2r(d+1+z)\beta\right)\frac{f'}{f}\nb\\
&&+2r(z\beta+d\gamma_1)\frac{g'}{g}+\left(2\Lambda  +\frac{q^2}{2g_e^2r^{2d}}\right)g^2\nb\\
&&+r^2\beta\frac{f'^2}{f^2}-2dz\beta-z^2\beta-d(d+1)\gamma_1=0,\\
\lb{feqB}
&&\frac{r^2\beta}{2}\frac{f'^2}{f^2}+(z\beta+d\gamma_1)r\frac{f'}{f}+\frac{z^2\beta}{2}+dz\gamma_1\nb\\
&&-\left(\Lambda  +\frac{q^2}{4g_e^2r^{2d}}\right)g^2+\frac{d}{2}(d-1)\gamma_1=0,~~~
 \eqn
 where
\bq
\lb{3.18}
 \Lambda\equiv \frac{\gamma_0\zeta^2}{2}.
\eq

To solve the above equations, we first note that Eq.(\ref{feqB}) can be   cast in the form,
 \bqn
 \lb{dyn1}
2d\beta\gamma_1W+\beta^2W^2-d^2\gamma_1^2(r_*^2-1)=0,
 \eqn
where
 \bqn
\lb{ff}
r_*(r)&\equiv&\sqrt{r_{s}^2 +\frac{2\beta g^2}{d^2\gamma_1^2}\left(\Lambda +\frac{2q^2}{8g_e^2r^{2d}}\right)},\nb\\
W(r) &\equiv& z + r\frac{f'(r)}{f(r)}, \nb\\
r_{s}^2 &\equiv& 1-\frac{(d-1)\beta}{d\gamma_1}.
 \eqn
Inversely, we find that
 \bq
\lb{2.10}
g^2 =\frac{d^2\gamma_1^2}{2\beta}\left(r_*^2-r_s^2\right)\left(\Lambda +\frac{2q^2}{8g_e^2r^{2d}}\right)^{-1}.
 \eq
From Eq.(\ref{dyn1}), we obtain
\bq
\lb{Fpm}
W=s\frac{1+\epsilon  r_*(r)}{1-s},
\eq
where $\epsilon =\pm1$, and
\bq
\lb{Fpm0}
 s\equiv\frac{d\gamma_1}{d\gamma_1-\beta}.
\eq
Then, from the stability conditions (\ref{2.20b}), we get
 \bq
\lb{s0} 1\le s<\frac{d-1}{d-2},
 \eq
which implies that
\bq
\lb{s0A}
0<r_s\le1,
\eq
where the equality holds only when $\beta = 0$ or $s=1$.

When  $s\not=1$, inserting Eq.(\ref{Fpm})  into the Hamiltonian constraint,
we obtain a master equation for $r_*(r)$,
 \bqn
\lb{hami3r0}
\nb\\
\frac{2dq^2r_*}{2q^2+C_2r^{2d}}+\frac{rsr_*^2r_*'}{1-d+(d-2)s+sr_*^2}\nb\\
=\epsilon \left(\frac{1}{1-s}-2+d\right) + \left(d+\frac{s}{1-s}\right)r_*+rr_*',
 \eqn
 which can be further  rewritten as
 \bqn
\lb{hami3r}
(s-1)rr'_*+\Delta(r)\left( r^2_*-r_s^2\right)
\left(r_*+\epsilon  {\cal{D}}(r)\right)=0,
 \eqn
where $\Delta = {s}/{\cal D}$, and
 \bqn
 \lb{pms}
{\cal{D}} &\equiv& \frac{[(d-1) + (2-d)s][2q^2+C_2r^{2d}]}{2q^2s+ C_2r^{2d}[d + (1-d)s]},\nb\\
 C_2&\equiv& 8g_e^2\Lambda.
 \eqn

When $q=0$,   we find that
 \bqn
\lb{pms2}
{\cal{D}} &=&
\frac{d\gamma_1-(d-1)\beta}{d(\gamma_1-\beta)},\; (q = 0),
 \eqn
which reduces to  the case   considered in \cite{SLWW,LSWW}. So, in
the rest of the paper we consider only the case $q\not= 0$. Unlike
the case without the electromagnetic field, now it is found
difficult to find the general solutions of Eq.(\ref{hami3r}).
Therefore, in the following we consider some particular cases.

\subsection{$s=1$}

When $s =1$, from Eq.(\ref{Fpm0}) we find that  this implies  $\beta =
0$. Then, the stability and ghost-free conditions require
\bq
\lb{aa}
\lambda = 1, \; (\beta = 0)
\eq
 as one can see from Eq.(\ref{2.17}). Therefore,
whenever we consider the case $s = 1$ (or $\beta = 0$) we always assume that $\lambda
= 1$. To study the solutions further, we consider  the two cases $d \not= 1$ and $d = 1$, separately.

\subsubsection{$d\ge 2$}

Then, we find  Eqs.(\ref{feqA}) and (\ref{feqB}) reduce to
\bqn
\lb{feqAb}
&&2rd\gamma_1\frac{g'}{g}+\left(2\Lambda+\frac{q^2}{2g_e^2r^{2d}}\right)g^2 \nb\\
&& ~~~~~~~~~~~~ -d(d+1)\gamma_1=0, \\
\lb{feqBb}
&&d\gamma_1r\frac{f'}{f}+dz\gamma_1+\frac{d}{2}(d-1)\gamma_1 \nb\\
&& ~~~~~~~~~~~-\left(\Lambda +\frac{q^2}{4g_e^2r^{2d}}\right)g^2=0,
\eqn
from which  we get
\bqn
\lb{feqC} g^2&=&4g_e^2d(d^2-1)\gamma_1
r^{2d+1}\big[(d-1)r^d(C_2r^{d+1}\nb\\
&&+4g_0g_e^2d(d+1)\gamma_1)-2q^2(d+1)r\big]^{-1}, \nb\\
f&=&f_0g^{-1}r^{1-z},
\eqn
where $f_0$ and $g_0$ are two integration constants. Rescaling $t \rightarrow f_0^{-1}t$, the metric can be cast in the form,
\bqn
\lb{feqD3} ds^2=-F(r)dt^2+\frac{dr^2}{F(r)}
+r^2\delta_{ij}dx^idx^j,~~~~
\eqn
where
\bq
\lb{feqD4} F(r) = - \frac{2m}{r^{d-1}}+ \frac{Q^2}{r^{2d-2}}-\frac{2\Lambda_{e} r^2}{d(d+1)},
\eq
with $g_0 \equiv -2m$, and
\bq
\lb{feqD5} Q^2 \equiv \frac{q^2}{2|\gamma_1|g_e^2d(d-1)},\;\;\; \Lambda_e \equiv \frac{\Lambda }{|\gamma_1|}.
\eq
Note that in writing the above expressions we had used the fact
\bq
\lb{gammaA}
\gamma_1 < 0, \; (d \ge 2)
\eq
 as one can see from Eq.(\ref{2.20c}).
The corresponding Ricci scalar of the surfaces $t= $ Constant  is given by
 \bqn
 \lb{feqD6}
R=-\frac{r^{2d}C_2+2q^2}{4g_e^2\gamma_1r^{2d}},
 \eqn
which is singular at $r=0$ when $q\not=0$.  On the other hand, the $(d+2)$-dimensional Ricci scalar is given by
 \bqn
 \lb{feqD6a}
{}^{(d+2)}R&=&
-F''-2d\frac{F'}{r}-d(d-1)\frac{F}{r^2}\nb\\
&=&2\frac{d+2}{d}\Lambda_e-(d-2)(d-1)\frac{q^2}{r^{2d}},
 \eqn
 which is also only singular at $r = 0$, provided that  $q \not=0$.

The above solutions are nothing
but the topologically Reissner-Nordstrom (anti-) de Sitter   solutions, and the Penrose diagrams have been given for various possibilities for
$d = 2$ in  \cite{Yumei}. It can be shown that these diagrams can be easily generalized  to the case with any $d$.

\subsubsection{$d=1$}

In this case,  since $\lambda = 1$,  from Eq.(\ref{2.20g}) we find that
\bq
\lb{BB}
\gamma_1\;\; {\mbox{is arbitrary}},\; (d = 1)
\eq
 and can take any of  real values.
Thus, from Eqs.(\ref{feqA}) and (\ref{feqB}) we find that

\bqn
\lb{s0a}
f&=&f_0r^{-z}\sqrt{\left|C_2r^2+8g_e^2\gamma_1g_0+4q^2\ln(r)\right|},\nb\\
g^2&=&\frac{8g_e^2\gamma_1r^2}{C_2r^2+8g_e^2\gamma_1g_0+4q^2\ln(r)}.
\eqn
Then, after rescaling the coordinate $t$,  the metric can be cast in the form,
\bqn
\lb{3.23a}
ds^2=-{{F}}(r)dt^2+\frac{dr^2}{{{F}}(r)}  +r^2dx^2,~~~~
\eqn
but now with
\bq
\lb{3.23}
{{F}}(r) =   -M  + A  \ln(r) + B r^2,
\eq
where
\bq
\lb{3.24}
M \equiv -g_0,\;\;\;
{{A}} \equiv \frac{q^2}{2 \gamma_1g_e^2},\;\;\;
{{B}} \equiv \frac{ \Lambda}{ \gamma_1}.
\eq
The corresponding 2d Ricci scalar is given by
 \bqn
 \lb{s0d}
R=-\frac{{{A}} + 2{{B}} r^2}{r^2},
 \eqn
which is singular only at $r = 0$ for $A \not= 0$. The nature of this singularity depends on the signs  of $B$. In particular, when $B > 0$ and $A < 0$, the above solutions are identical
to the charged Banados-Teitelboim-Zanelli (BTZ) black hole \cite{BTZ} with
\bq
\lb{3.39}
\left(M, Q, \Lambda_e\right) = \left(- g_0, \sqrt{-A}, - B\right),
\eq
where $M, \; Q$ and  $\Lambda_e$ denotes, respectively, the  BTZ black hole mass, charge, and  the  the effective cosmological constant.

But, the  solutions   of Eq.s(\ref{3.23}) are more general than the charged BTZ black hole. In particular, $A$ is not necessarily negative, as now $\gamma_1$ is a free parameter. To study
 these solutions further, we consider the cases $B > 0$, $B = 0$ and $B < 0$, separately.

{\bf Case i) $B >  0$.} In this case, one can see that the space-time is always asymptotically anti-de Sitter. However, depending on the signs of $A$, the solutions can have different properties.
To see these clearly, let us consider the cases $A > 0$, $A = 0$ and $A < 0$, separately.

\begin{figure}[tbp]
\centering
\includegraphics[width=8cm]{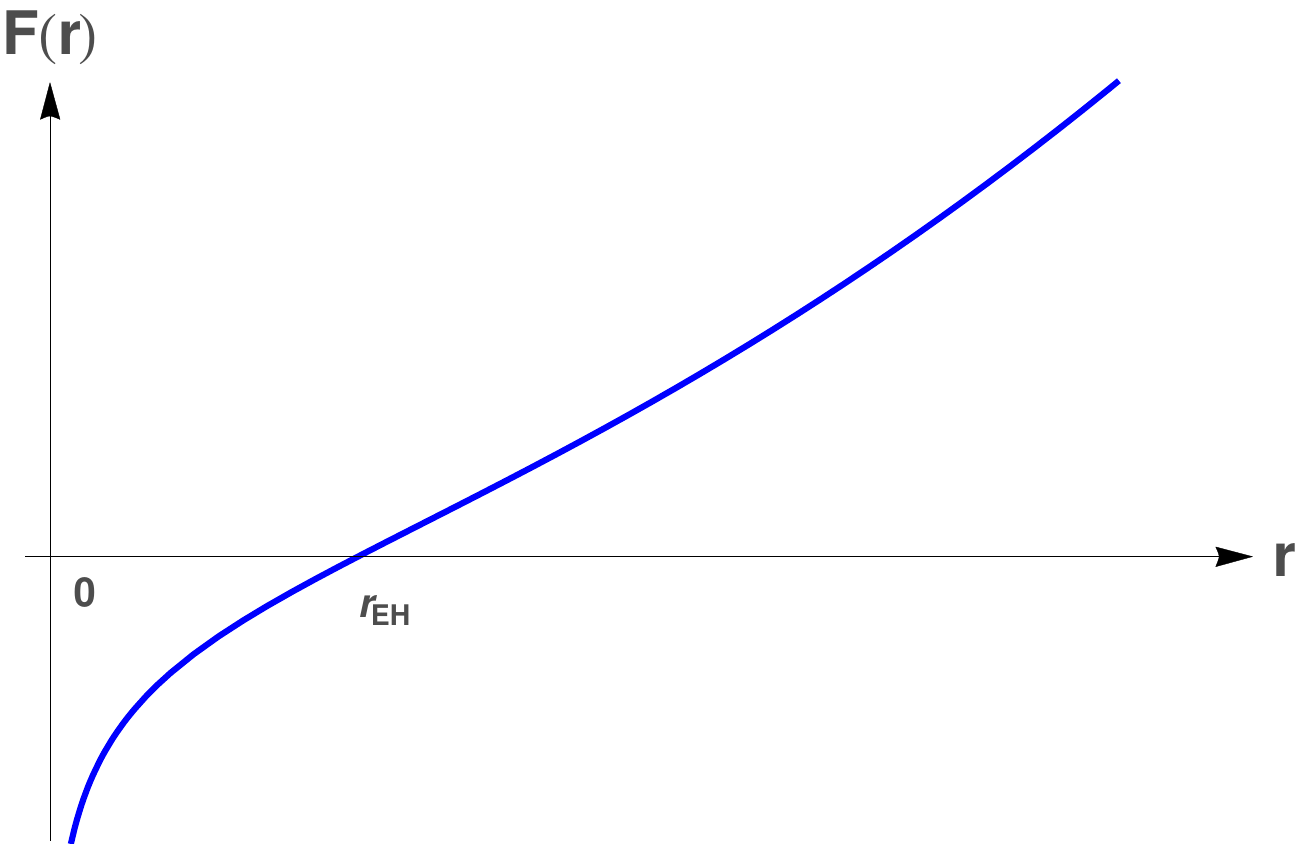}
\caption{The function ${{F}}(r)$ defined in Eq.(\ref{3.23})  vs $r$ for $B > 0$ and $A > 0$, where $r = r_{EH}$ represents the Killing (event) horizon of the black hole. The space-time is singular at $r=0$.  }
\label{figA}
\end{figure}

{\bf Case i.1) $B > 0, \; A >  0$:} In this case, we find that
\bq
\lb{3.26a0}
\gamma_1 > 0,\; \Lambda > 0,
\eq
for which we have
\bq
\lb{3.26}
{{F}}(r) =
\cases{
-\infty, & $r = 0$,\cr
< 0, & $r < r_{EH}$,\cr
0, & $r = r_{EH}$,\cr
> 0, & $r > r_{EH}$,\cr
\infty, & $r = \infty$,\cr
}
\eq
where $r_{EH}$ is the unique real root of ${{F}}(r) = 0$, as shown in Fig. \ref{figA}.
Thus, in this case the singularity is covered by the Killing horizon located at $r = r_{EH}$.
The space-time  is asymptotically anti-de Sitter with the effective cosmological constant  given by
$\Lambda_{\mbox{\small eff} }= - B$.
The  surface gravity on the Killing horizon is given by
\bq
\kappa_{EH} = \frac{B(r_{EH}^2 +  r_c^2)}{r_{EH}},
\eq
which is always positive, where $r_c \equiv \sqrt{A/(2B)}$.

{\bf Case i.2) $B > 0, \; A =  0$:} In this case, we have $q = 0$, and the corresponding solutions are not charged. Then,   we find, 
\bqn
{{F}}(r)  = - M + B r^2 = \cases{-M, & $ r = 0$,\cr
 0, & $ r = r_{EH}$,\cr
 \infty, & $ r = \infty$,\cr}
\eqn
where $r_{EH} \equiv \sqrt{M/B}$. The surface gravity on the Killing horizon now is given by
\bq
\kappa_{EH} = \sqrt{BM}.
\eq

\begin{figure}[tbp]
\centering
\includegraphics[width=8cm]{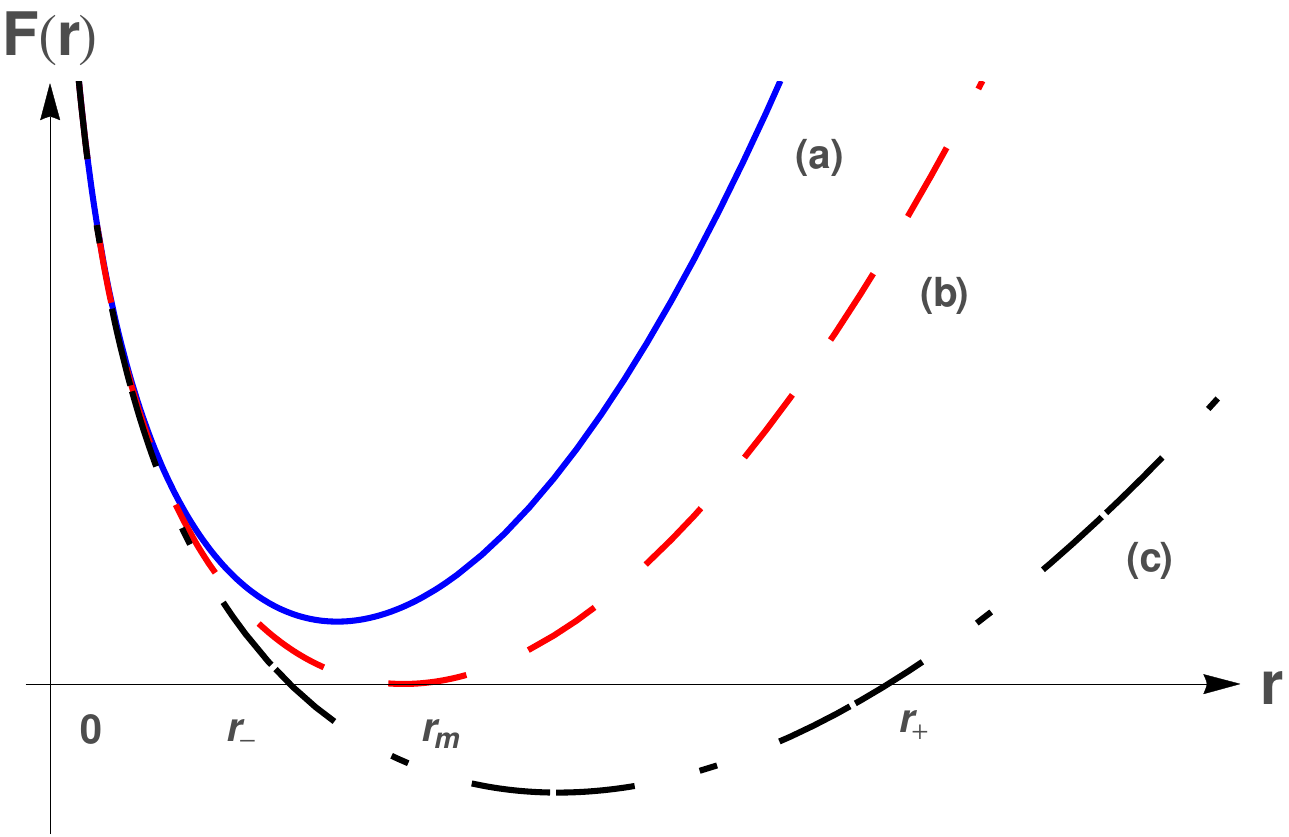}
\caption{The function ${{F}}(r)$ defined in Eq.(\ref{3.23})  vs $r$ for $B > 0$ and $A < 0$:
(a) $\Lambda < \Lambda_c$; (b)   $\Lambda = \Lambda_c$; and
(c)  $0 > \Lambda > \Lambda_c$, where $\Lambda_c$ is defined in Eq.(\ref{3.26c}). }
\label{figA2}
\end{figure}

{\bf Case i.3) $B > 0, \; A <  0$:} In this case, we find
\bq
\lb{3.26a}
\gamma_1 <0, \;\;\; \Lambda < 0,
\eq
and $F(r) \rightarrow \infty$ for both $r \rightarrow 0$ and $r \rightarrow \infty$. Thus, depending on the values of $B$ (or $\Lambda$), the equation $F(r) = 0$ can have two, one or none real roots.
In particular, we have $F'(r) = -2B(r^2 - r_m^2)/r$, where $r_m \equiv \sqrt{|A|/(2B)}$. Thus, we find
\bqn
\lb{3.26b}
F(r_m) &=& \frac{|A|}{2}\left[\ln\left(\frac{2B}{|A|}\right) + 1\right] - M \nb\\
&=&  \cases{ > 0,  & $ \Lambda <  \Lambda_c$,\cr
 = 0, & $\Lambda = \Lambda_c$,\cr
 < 0, & $ 0 > \Lambda >  \Lambda_c $,\cr}
\eqn
where
\bq
\lb{3.26c}
\Lambda_c \equiv - \frac{1}{2} |\gamma_1 A| e^{\frac{2M - |A|}{|A|}}.
\eq
Then, the behavior of $F(r)$ is illustrated in Fig.\ref{figA2}, from which we can see that the singularity at $r = 0$ is naked in Case (a), in which we have $\Lambda < \Lambda_c$.
In case (c), where $0 >  \Lambda > \Lambda_c$, there exist two Killing horizons, located, respectively, at $r = r_{-}$ and $r = r_{+}$. In this case, the Penrose diagram is similar to the
Reissner-Nordstrom anti-de Sitter solutions. In particular, the surface gravity
  is negative at $r = r_{-}$, while  positive at  $r = r_{+}$, as one can
see from Fig. \ref{figA2}. In Case (b), the two horizons become degenerate, and the surface gravity is zero.

\begin{figure}[tbp]
\centering
\includegraphics[width=8cm]{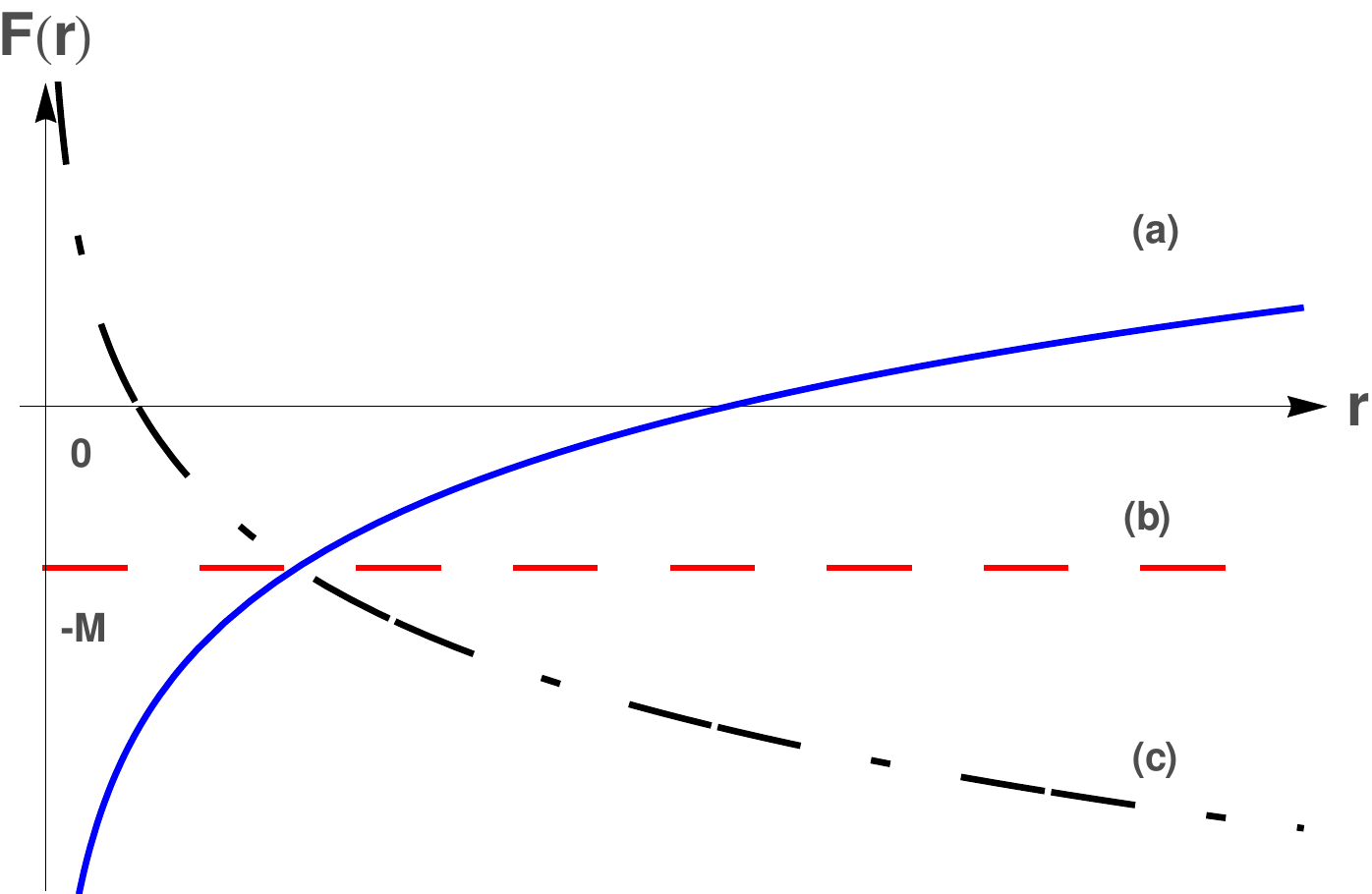}
\caption{The function ${{F}}(r)$ defined in Eq.(\ref{3.23})  vs $r$ for $B = 0$:
(a) $A > 0$; (b)   $A = 0$; and
(c)  $ A < 0$. }
\label{figA3}
\end{figure}

{\bf Case ii) $B  = 0$.} In this case, we find that
\bq
\lb{3.26d}
F(r) = - M + A \ln(r).
\eq
Thus, depending on the values of $A$, the solutions can have different properties. In particular, when $A >0$ ($A < 0$)  the function $F(r)$ is monotonically increasing (deceasing)
as shown in Fig. \ref{figA3}. Then,  there always exists a point $r_{EH}$ at which $F(r_{EH}) = 0$.  The surface gravity on this Killing horizon is positive for $A >0$ and negative for
$A < 0$. When $A = 0$, the space-time is flat.

\begin{figure}[tbp]
\centering
\includegraphics[width=8cm]{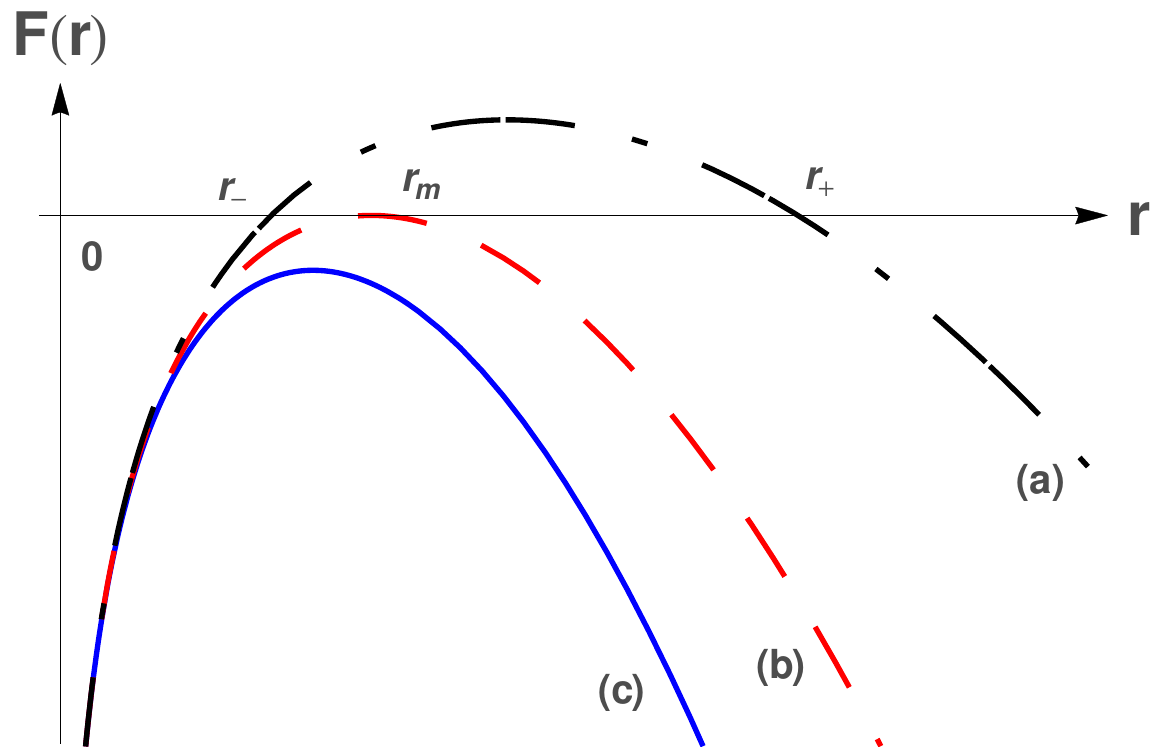}
\caption{The function ${{F}}(r)$ defined in Eq.(\ref{3.23})  vs $r$ for $B < 0$ and $A > 0$. (a) $|B| < B_m$;  (b) $|B| = B_m$; and (c) $|B| > B_m$;  where $B_m$ is defined
in Eq.(\ref{3.28}).  }
\label{figB}
\end{figure}

{\bf Case iii) $B <  0$:}  In this case, we have
\bqn
\lb{3.27}
{{F}}'(r)  &=& 2|B|\frac{r_m^2-r^2}{r},\nb\\
{{F}}(r_m) &=& -M + \frac{A}{2}\left[\ln\left(\frac{A}{2|B|}\right) -1\right] \nb\\
&=& \cases{ > 0, & $|B| < B_m$,\cr
= 0, & $|B| = B_m$,\cr
< 0, & $|B| > B_m$,\cr}
\eqn
where
\bq
\lb{3.28}
r_m \equiv \sqrt{\frac{A}{2|B|}},\;\;\;
B_m \equiv \frac{1}{2}A e^{\frac{-(2M+A)}{A}}.
\eq
Fig. \ref{figB} shows the curve of ${{F}}(r)$ vs $r$. In the case $|B| > B_m$,
 we can see that ${{F}}(r)$ is always negative, and the coordinate $r$ is timelike, and the corresponding space-time is dynamical, and has a spacelike naked singularity at $r= 0$.

In the case $|B| = B_m$, a coordinate singularity appears at $r = r_m$, as it can be seen from Eq.(\ref{s0d}), which now takes the form,
\bqn
\lb{3.29}
R=-2|B|\frac{r_m^2 - r^2}{r^2}.
\eqn
Therefore, in order to obtain a geodesically maximal space-time, extension across this surface is needed.   Such an extension is quite  similar to the one of the extremal  case, $9 \Lambda m^2 =1$,  of the
Schwarzschild-de Sitter solution, $F(r) = 1 - 2m/r - \Lambda r^2/3$. The corresponding  surface gravity is zero, as it can be seem from
Eq.(\ref{3.27}).

In the case $|B| < B_m$, ${{F}}(r)$ is positive only for $ r_- < r < r_+$, where  $r_{\pm}$ are the two real roots of   ${{F}}(r) = 0$, with $r_{+} > r_{-}$. From Eq.(\ref{3.29}) we can see that the singularities at
$r= r_{\pm}$ represent horizons, and the extensions beyond  these horizons are similar to the Schwarzschild--de Sitter
 solution with $0< \Lambda < 1/(9m^{2})$. The corresponding surface gravity is given by
 \bq
 \lb{3.29a}
 \kappa_{\pm} =  |B|\frac{r_m^2-r_{\pm}^2}{r_{\pm}},
 \eq
 which is positive only when $r_m > r_{\pm}$.

It should be noted that  the above analysis holds only  for  $A > 0$.

When $A = 0$, we have $F(r) = - M - |B|r^2 < 0$, and the solutions represent dynamical space-time,
in which $r$ ($t$) is always timelike (spacelike),
 and the singularity at $r = 0$ is naked.

 When $A < 0$, we find that
 \bqn
 \lb{3.27aa}
 F(r) &=& - M - |A|\ln(r) - |B|r^2 \nb\\
 &=& \cases{\infty, & $r = 0$,\cr
 - \infty, & $r = \infty$,\cr}
 \eqn
 which is monotonically decreasing function of $r$ [cf. Curve (c) in Fig.\ref{figA3}], and asymptotically approaches to the de Sitter space-time.
 The surface gravity at $r = r_{EH}$ now clearly is negative.

 \subsection{$\Lambda =0$}

\begin{figure}[tbp]
\centering
\includegraphics[width=8cm]{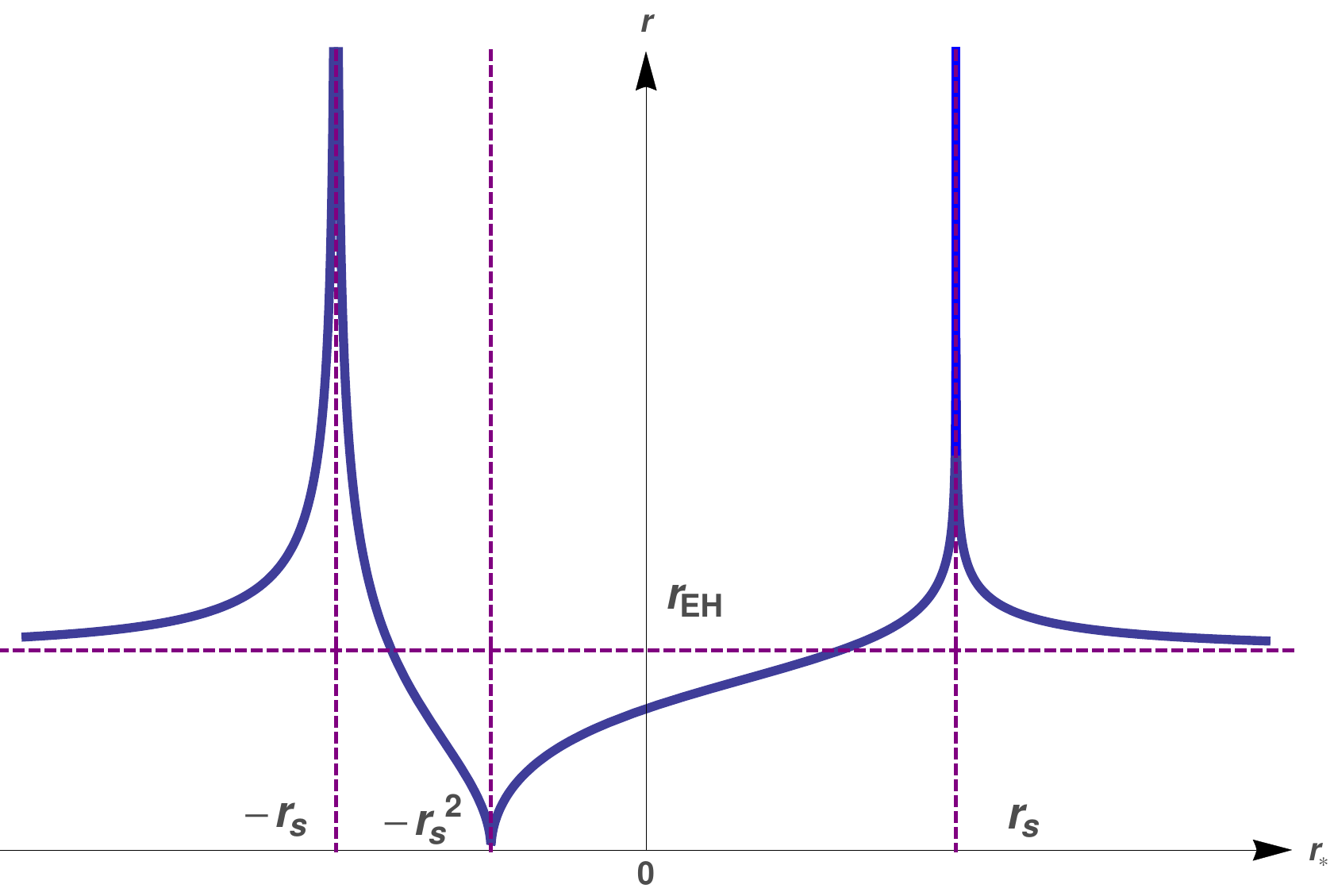}
\caption{The function $r\equiv r (r_*)$ defined in Eq.(\ref{case2res1})  for $1 < s < \frac{d-1}{d-2}$.}
\label{fig1}
\end{figure}

In this case, from Eq.(\ref{pms}) we find
\bq
\lb{3.30} {\cal{D}}=r_s^2=\frac{d-1+s(2-d)}{s}.
\eq
Since $s \not=1$, from Eqs.(\ref{s0}) and(\ref{s0A}) we find that the stability condition (\ref{2.20b}) leads to
\bq
\lb{3.30a}
0<r_s < 1,\;\;\; 1< s<\frac{d-1}{d-2}.
\eq
Without loss of the generality, we set $\epsilon =1$ in Eq.(\ref{Fpm}).
Then,  Eq.(\ref{hami3r}) can be cast in the form,
 \bqn
\lb{case2hami1}
\frac{dr}{r}&=&\Bigg[\frac{r_s-1}{2(d-1)\left(r_*-r_s\right)}-\frac{r_s+1}{2(d-1)\left(r_*+r_s\right)}\nb\\
&&+\frac{1}{(d-1)\left(r_*+r_s^2\right)}\Bigg]dr_*,
 \eqn
from which we find that the general solution,
 \bqn
  \lb{case2res1}
r(r_*)=r_{EH}\frac{\left|r_*+r_s^2\right|^{\frac{1}{d-1}}}{\left|r_*-r_s\right|^{\frac{1-r_s}{2(d-1)}}\left|r_*+r_s\right|^{\frac{r_s+1}{2(d-1)}}},
 \eqn
where $r_{EH}$ is a constant. Then, we obtain
 \bqn
\lb{scase2} r(r_*) = \cases{r_{EH} , & $r_* \rightarrow -\infty$,\cr
\infty, & $r_*  = -r_s$,\cr 0, & $r_*  =  - r_s^{2}$,\cr \infty, &
$r_* = r_s$,\cr r_{EH} , & $r_* \rightarrow + \infty$,\cr}
 \eqn
Fig. \ref{fig1} shows the curve of $r(r_*)$ vs $r_*$.

On the other hand, from Eqs.(\ref{ff}) and (\ref{Fpm}) we find
\bqn
\lb{case2f1a}
\frac{df}{f}&=&\left[\frac{d-1+z-zr_s}{2(d-1)\left(r_*-r_s\right)}+\frac{d-1+z+zr_s}{2(d-1)\left(r_*+r_s\right)}\right.\nb\\
&&\left. +\frac{1-d-z}{(d-1)\left(r_*+r_s^2\right)}\right]dr_*,
\eqn
which has the general solution,
\bqn
\lb{case2f1c}
f=f_0\left|r_*-r_s\right|^{\delta_1}\left|r_*+r_s\right|^{\delta_2}\left|r_*+r_s^2\right|^{-\frac{d-1+z}{d-1}},
~~~~
\eqn
where $f_0$ is another integration
constant, and
\bq
\lb{2.20} \delta_1 \equiv \frac{d-1+z-zr_s}{2(d-1)},\;\;\; \delta_2
\equiv \frac{d-1+z+zr_s}{2(d-1)}.
\eq
Finally, from Eq.(\ref{2.10}) we find
\bqn
\lb{case2g} g^2(r_*)=C_1r^{2d}\left(r_s^2-r_*^2\right),
\eqn
where $C_1 \equiv - 2d^2{g_e^2\gamma_1^2}/{(q^2\beta)}$. Since $\beta < 0$, we find that  $C_1$ is always positive, $C_1 > 0$. Then, to have $g^2$ positive, we must restrict ourselves
to the region $ | r_*| < r_s$. Hence,  the corresponding metric takes the form,
\bq
\lb{case2ga} ds^2 =-N^2dt^2+G^2dr^2+\delta_{ij}r^2dx^idx^j,
\eq
where
\bqn
\lb{case2i} N^2&=&N_0^2 \frac{r_s^2-r_*^2}{\left(r_*+r_s^2\right)^2}, \nb\\
G^2&=&C_1 r^{2(d-1)}\left(r_s^2 - r_*^2\right),
\eqn
where $N_0 \equiv r_{EH}^{z}f_0$.
The corresponding Ricci scalar  is given by
\bqn
\lb{RS}
R&=&R_0\frac{(r_*-r_s)^{\frac{1-dr_s}{d-1}}(r_*+r_s)^{\frac{1+dr_s}{d-1}}}{(r_*+r_s^2)^{\frac{2d}{d-1}}}\nb\\
&&\times\left[r_s^4+2r_*^2+r_s^2(2r_*-1)\right],
\eqn
where
\bqn
\lb{RS0} R_0 &\equiv&
\frac{d(d-1) }{r_{EH}^{2d}C_1(1-r_s^2)r_s^2}.
\eqn
Thus, the space-time is always  singular at  $r_* =  -r_s^2$, which divides the regions $ | r_*| < r_s$ into two, $-r_s < r_* < - r_s^2$ and $-r_s^2 < r_* < r_s$.

In the region $-r_s < r_* < -r_s^2$, we have $r \in (0, \infty)$. So, in this region  the space-time is already complete with a naked singularity located at $r_* = -r_s^2$ (or $r = 0$). As
$r_* \rightarrow - r_s^{-}$ or $r \rightarrow \infty$, we find that the metric (\ref{case2ga}) takes the asymptotical form,
\bqn
\lb{3.33a}
 ds^2&\simeq&L^2 \bar{r}^{-\frac{2(d-\theta)}{d}}\left(- \bar{r}^{-2(z-1)}d\bar{t}^2 + d\bar{r}^2  + \delta_{ij}dx^idx^j\right),\nb\\
\eqn
with
\bq
\lb{3.57}
\theta = \frac{d(1+d r_s)}{(d-1)r_s},\;\;\;
z = \frac{d(1+r_s)}{(d-1)r_s}.
\eq
The above metric represents  Lifshitz space-times with hyperscaling violation.

In the region $-r_s^2 < r_* < r_s$, the space-time is also singular  at  $r_* =  r_s$ when $r_s>\frac{1}{d}$.   Then, the physical interpretation is unclear, as the space-time is singular at
both $r = 0$ and $r= \infty$. However, when $r_s \le \frac{1}{d}$, the space-time is free of this kind of singularity. In fact, the space-time  also take asymptotically the form (\ref{3.30a})
but now with
\bq
\lb{3.58}
\theta = \frac{d(d r_s -1)}{(d-1)r_s},\;\;\;
z = - \frac{d(1-r_s)}{(d-1)r_s}.
\eq

 \section{Universal Horizons and static charged black holes }
\renewcommand{\theequation}{4.\arabic{equation}} \setcounter{equation}{0}

According to the definition of the  universal horizons given in
Section II, they can exist only inside the Killing horizons. Then,
from the charged solutions presented in the last section, one can
see that this is the case only for the solutions given by
Eqs.(\ref{feqD3}) and (\ref{3.23a}), for which we have $(\beta,
\lambda) = (0, 1)$ \footnote{Universal horizons in (2+1)-dimensional
space-time with rotation was considered recently in \cite{SVV}, and
found that they exist also in the case $\beta = 0$. In this case, as
shown in Appendix A, the stability condition requires $\lambda =
1$.}.  Moreover, in order to keep $t$ being asymptotically timelike,
in the following  we shall consider only the case that the solutions
given by Eqs.(\ref{feqD3}) and (\ref{3.23a}) are asymptotically
anti-de Sitter, that is,
 \bq
\lb{4.0}
F(r) \simeq \left(\frac{r}{\ell_{d}}\right)^2,
 \eq
as $r \rightarrow \infty$.

To solve the khronon equation (\ref{eq1.8}) in general case is found
to be very difficult. But, when $c_1 + c_4 = 0$, Eq.(\ref{eq1.8})
has a simple solution $u^r = r_B/r^d$, where $r_{B}$ is an
integration constant. Then, from the condition
$u_{\lambda}u^{\lambda} = -1$  we can find $u^t$, so finally we
have,
 \bqn
\lb{UHB2}
u^{\mu}=\delta^{\mu}_{t}
\frac{\sqrt{r_B^2+r^{2d}F(r)}}{r^dF(r)}-\delta^{\mu}_{r}
\frac{r_B}{r^d}.
 \eqn
Hence, we arrive at
 \bqn
\lb{UHB3}
u_{\mu}\zeta^{\mu} &=&-\frac{\sqrt{r_B^2+r^{2d}F(r)}}{r^d}.
\eqn
To assure that the four-velocity of the  khronon is well-defined in the whole space-time, especially inside the universal horizon,  we  require that  \cite{BBM} \footnote{When
$c_1 +c_4 = 0$ the speed of the khronon becomes infinitely large \cite{EA,BBM}. Then, the sound horizon coincides with the universal horizon, and the regularities required at the sound horizon
become the ones at the universal horizon. In the present case, it can be shown that Eq.(\ref{CD2}) is the sufficient condition to assure that the sound horizon is regular.}
\bq
\lb{CD2}
\left. \frac{d}{dr}\left(u_{\mu}\zeta^{\mu}\right)^2\right|_{r= r_{UH}}=0,
\eq
 on the universal horizon. Then, from Eq.(\ref{UHB3}) we find that
 \bqn
\lb{UHB4a}
&& r_B^2 = -F(r_{UH})r_{UH}^{2d},\\
\lb{UHB4b}
&& 2dF(r_{UH})+r_{UH}F'(r_{UH})=0.
\eqn
The corresponding surface gravity \cite{CLMV} is given by,
   \bqn
\lb{UHB6}
\kappa_{UH}&\equiv & \frac{1}{2} u^{\alpha} D_{\alpha} \left(u_{\lambda} \zeta^{\lambda}\right) \nb\\
&=&   \left. \frac{r_B}{2\sqrt{2}r^d}\sqrt{G''\left(r\right)}\right|_{r = r_{UH}},
\eqn
where
\bq
\lb{UHB6a}
G(r) \equiv \frac{r_{B}^2}{r^{2d}} + F(r).
\eq

To study the universal horizons further, we need to consider the cases $d \ge 2$ and $d = 1$, separately.

 \subsubsection{  $ d \ge 2$}

In this case, from Eq.(\ref{UHB4a}) we find that
 \bqn
\lb{UHB5}
r_B^2=-Q^2r_{UH}^2+2mr_{UH}^{1+d}+\frac{2\Lambda_er_{UH}^{2+2d}}{d+d^2},
\eqn
where $r_{UH}$ is a real and positive root of Eq.(\ref{UHB4b}), which now takes the form,
\bqn
\lb{UHB5a}
2\left|\Lambda_e\right|r_{UH}^{d+1} + dQ^2r_{UH}^{1-d} - d(d+1)m =0.
\eqn
Note that in writing the above equation we assumed $\Lambda_{e} < 0$ in order for the solutions to be asymptotically anti-de Sitter.

\begin{figure}[tbp]
\centering
\includegraphics[width=8cm]{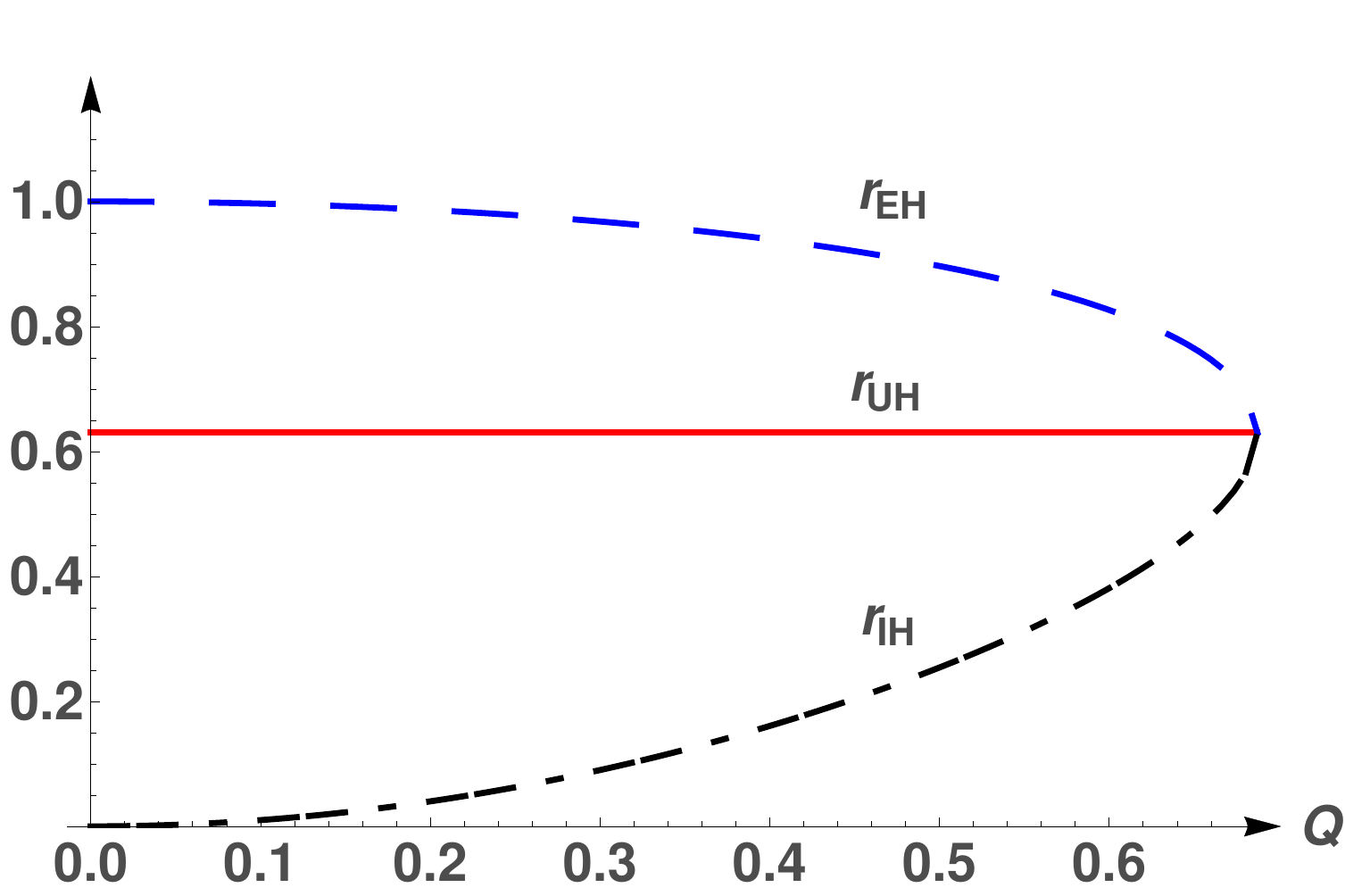}
\caption{The locations of the universal horizon $r = r_{UH}$, Killing (event) horizon $r = r_{EH}$, and  inner horizon $ r = r_{IH}$  for
 the solutions given by Eqs.(\ref{feqD3}) and (\ref{feqD4})  with $d=2,\; \Lambda_e=-\frac{d(d+1)}{2},\; m=\frac12$.}
 \label{h4}
\end{figure}

\begin{figure}[tbp]
\centering
\includegraphics[width=8cm]{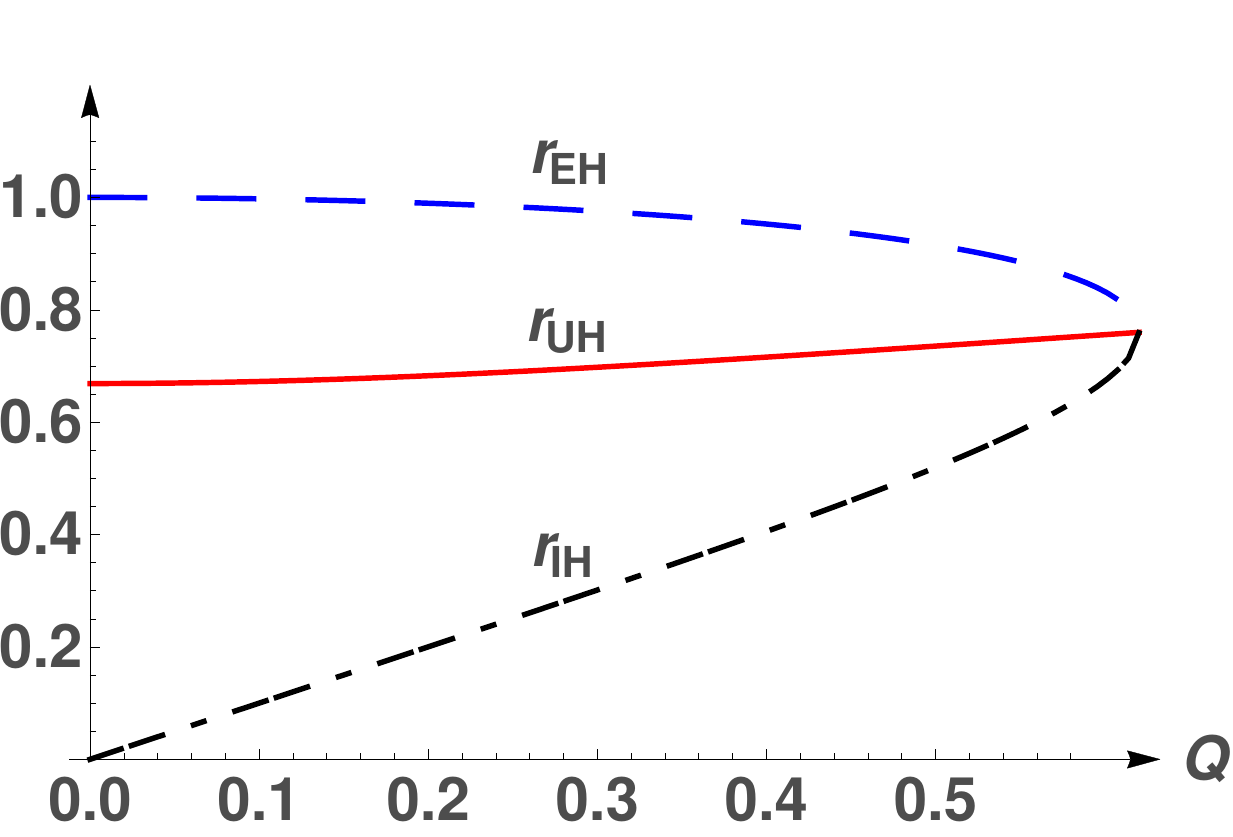}
\caption{The locations of the universal horizon $r = r_{UH}$, Killing (event) horizon $r = r_{EH}$ and  inner horizon $ r = r_{IH}$ for
 the solutions given by Eqs.(\ref{feqD3}) and (\ref{feqD4}) with $d=3$,  $\Lambda_e=-\frac{d(d+1)}{2}$ and $m=\frac12$.}
\label{h5}
\end{figure}

\begin{figure}[tbp]
\centering
\includegraphics[width=8cm]{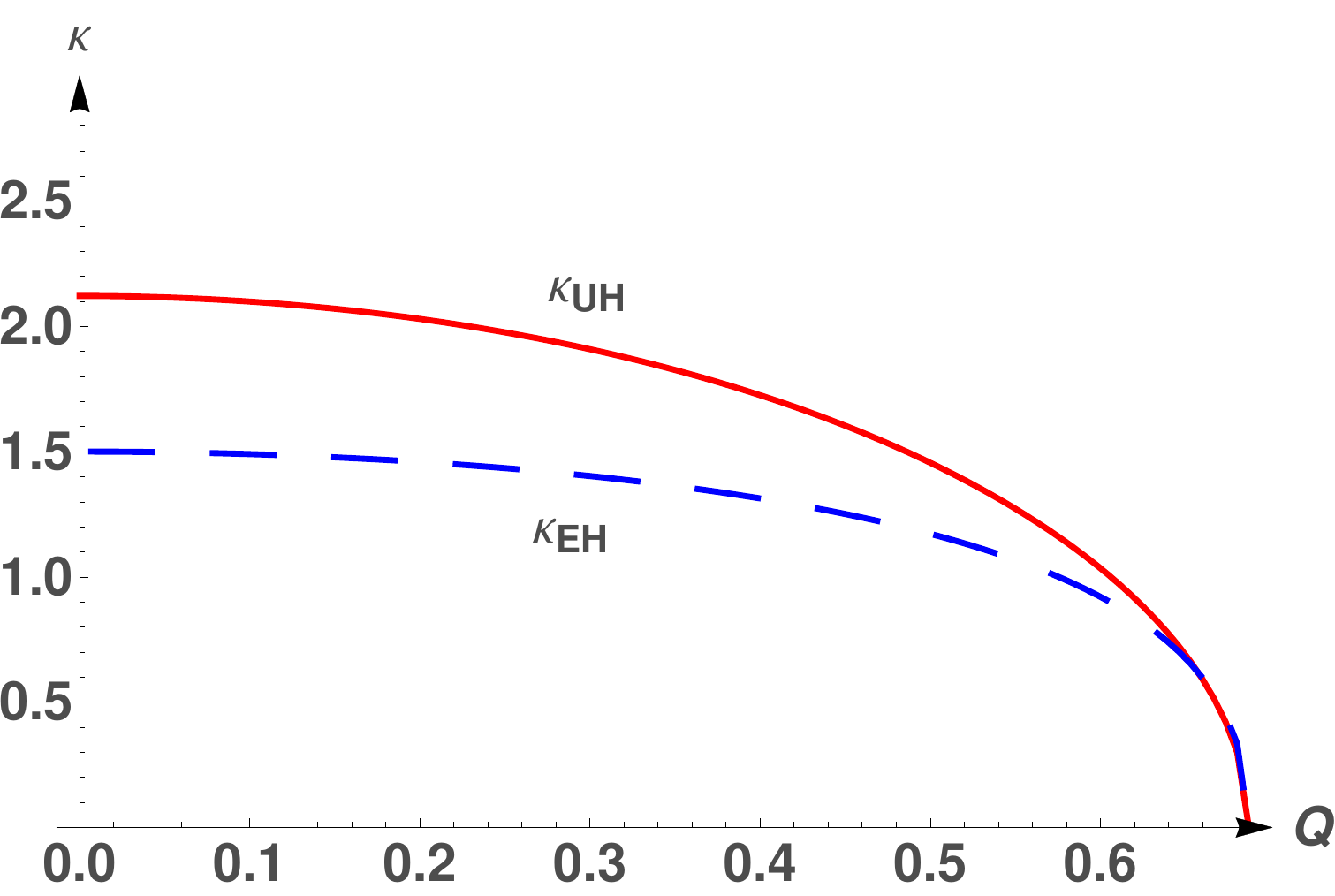}
\caption{The surface gravity $\kappa_{UH}$ on the universal horizon $r = r_{UH}$, and  the one $\kappa_{EH}$ on the event horizon  $r = r_{EH}$  for
 the solutions given by Eqs.(\ref{feqD3}) and (\ref{feqD4})  with $d=2,\;
\Lambda_e=-\frac{d(d+1)}{2},\; m=\frac12$.} \label{k4}
\end{figure}

\begin{figure}[tbp]
\centering
\includegraphics[width=8cm]{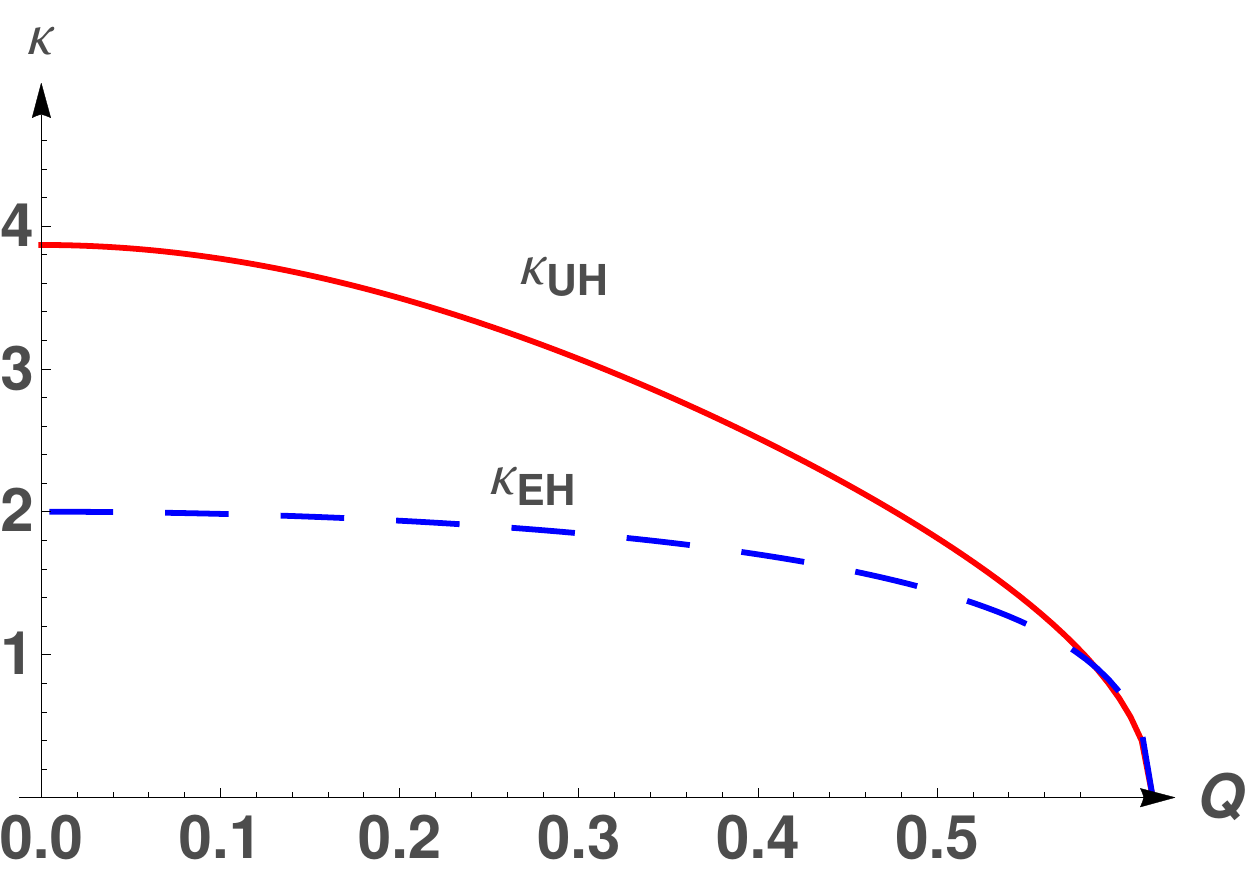}
\caption{The surface gravity $\kappa_{UH}$ on the universal horizon $r = r_{UH}$, and  the one $\kappa_{EH}$ on the event horizon
 $r = r_{EH}$  for the solutions given by Eqs.(\ref{feqD3})  and (\ref{feqD4}) with
 $d=3$, $\Lambda_e=-\frac{d(d+1)}{2}$ and $m=\frac12$.}
\label{k5}
\end{figure}

In Figs. \ref{h4} and \ref{h5}, we show the locations of the universal horizons vs the total charge $Q$ for $\Lambda_e=-\frac{d(d+1)}{2},\; m=\frac12$
 in the cases $d = 2$ and $d =3$. From these figures we can see that the universal horizons are always inside  the outer Killing horizon
$r = r_{+}$ but outside  the inner horizon $ r = r_{-}$. At the extremal  case, the universal horizon coincides with two Killing horizons, where the charge of black hole is given by 
\bq
\lb{cc}
Q_c^2=-2^{\frac{1-d}{1+d}}\frac{\Lambda_e}{d(d-1)}\left[\frac{-\Lambda_e}{M(d^2-1)}\right]^{-\frac{2d}{d+1}}.
\eq
When $Q^2 > Q_c^2$, the singularity located at the center $r = 0$ becomes naked, and no black hole exists.
On the other hand, from Figs. \ref{k4} and \ref{k5} we can see  that the surface gravity on the universal
horizons are always larger than those on the outer Killing horizons.

 \subsubsection{ $d = 1$ }

In this case the solutions are given by Eqs.(\ref{3.23a}) and (\ref{3.23}). Substituting $r_{B}$ given by Eq.(\ref{UHB4a}) into Eq.(\ref{UHB4b}) we find that
 \bqn
\lb{UHA5} 4Br_{UH}^2 + 2A\ln(r_{UH})+ (A-2M) =0.
\eqn
As shown in the last section, depending on the choice of the free parameter $B [= \Lambda/\gamma_1]$, the corresponding solutions have different properties.

\begin{figure}[tbp]
\centering
\includegraphics[width=8cm]{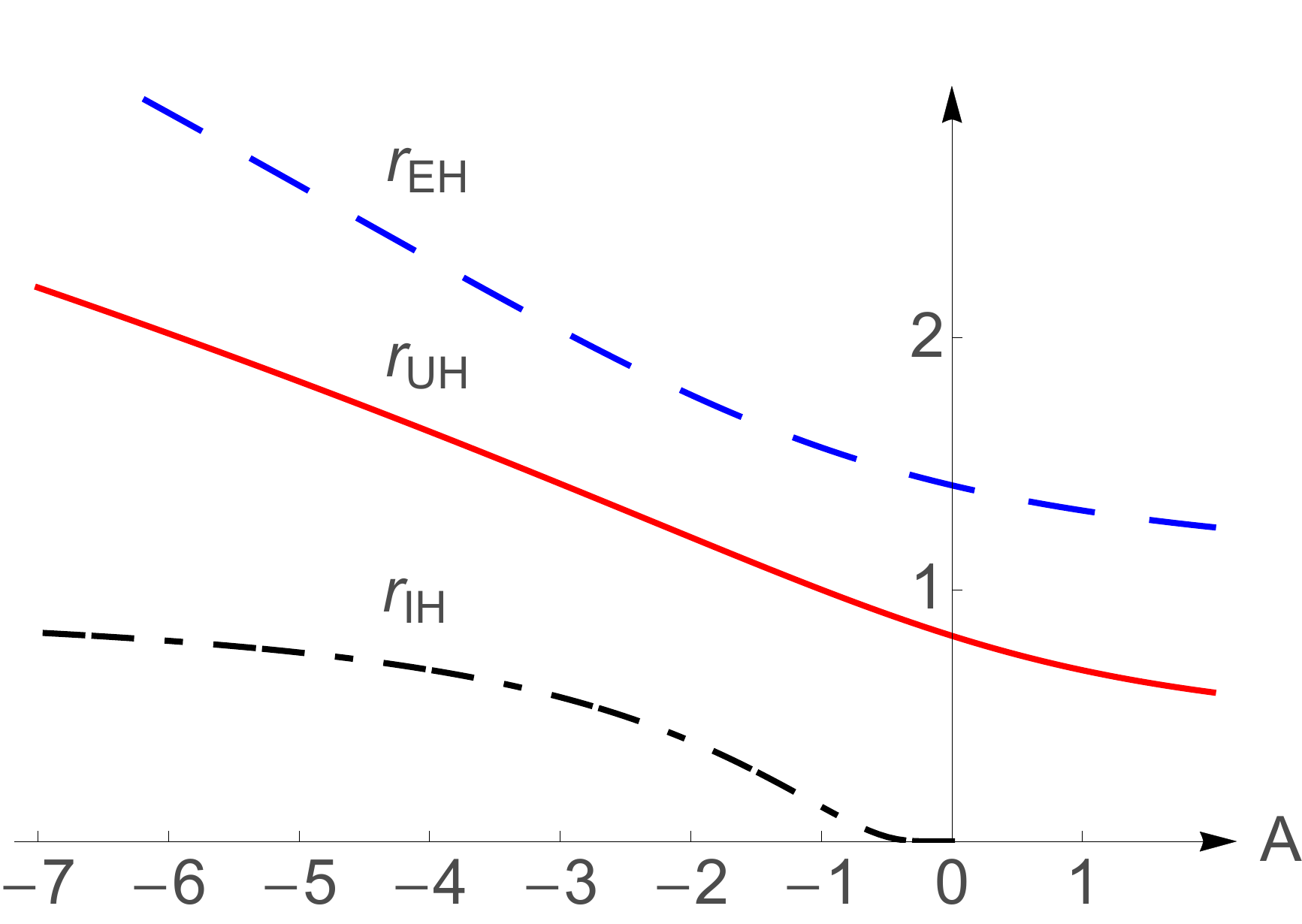}
\caption{The locations of the universal horizon $r = r_{UH}$, Killing (event) horizon $r = r_{EH}$, and  inner horizon $ r =
r_{IH}$  for  the solutions given by  Eqs.(\ref{3.23a}) and (\ref{3.23}) for $M > B > 0$.  When drawing this figure, we set $(M, B) = (2, 1)$.} \label{h3a}
\end{figure}

\begin{figure}[tbp]
\centering
\includegraphics[width=8cm]{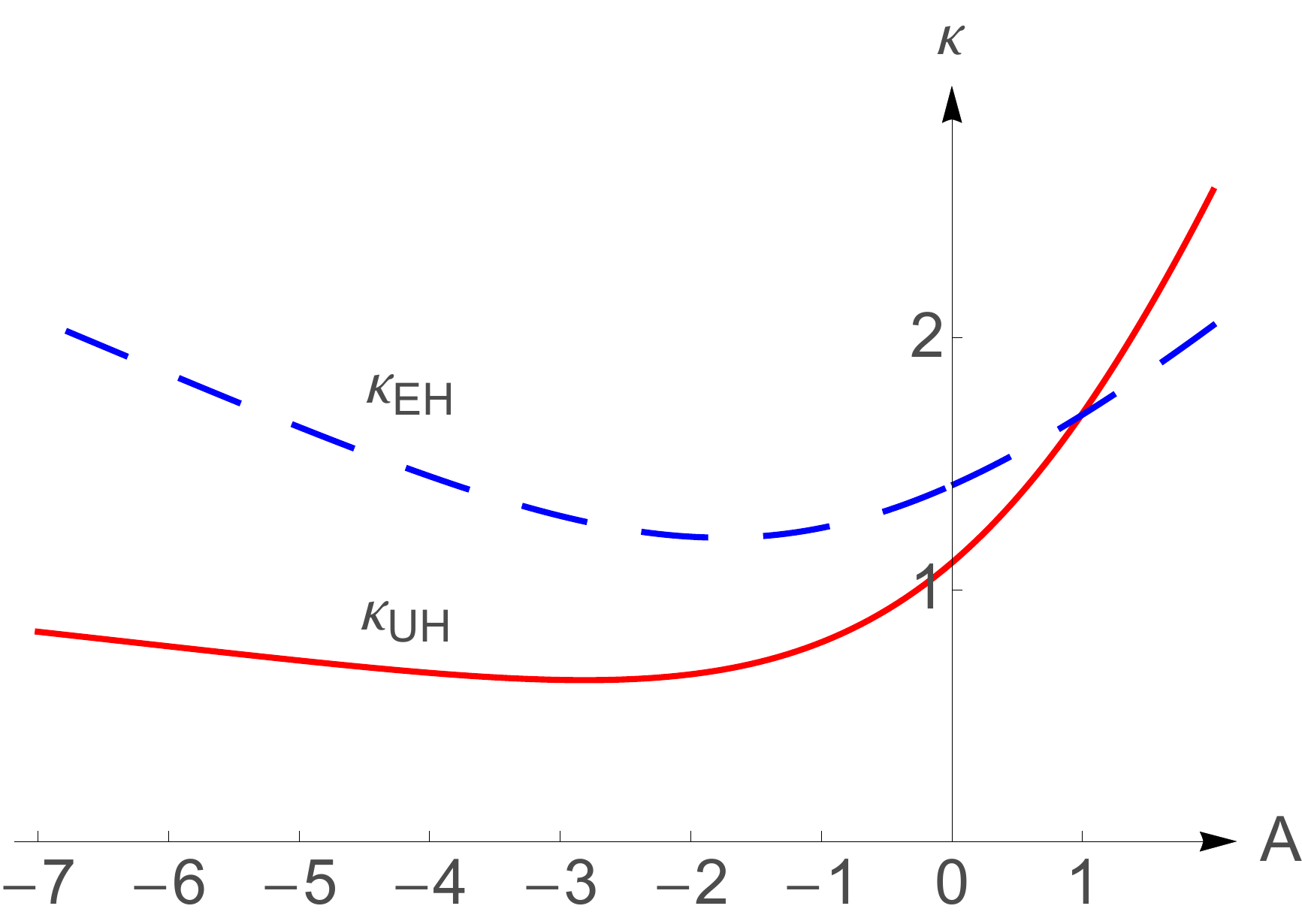}
\caption{The surface gravity $\kappa_{UH}$ on the universal horizon $r = r_{UH}$, and  the surface gravity $\kappa_{EH}$ on the event horizon
 $r = r_{EH}$  for the solutions given by Eqs.(\ref{3.23a}) and (\ref{3.23}) for $M > B > 0$. When drawing this figure, we set  $(M, B) = (2, 1)$.} \label{k3a}
\end{figure}

\begin{figure}[tbp]
\centering
\includegraphics[width=8cm]{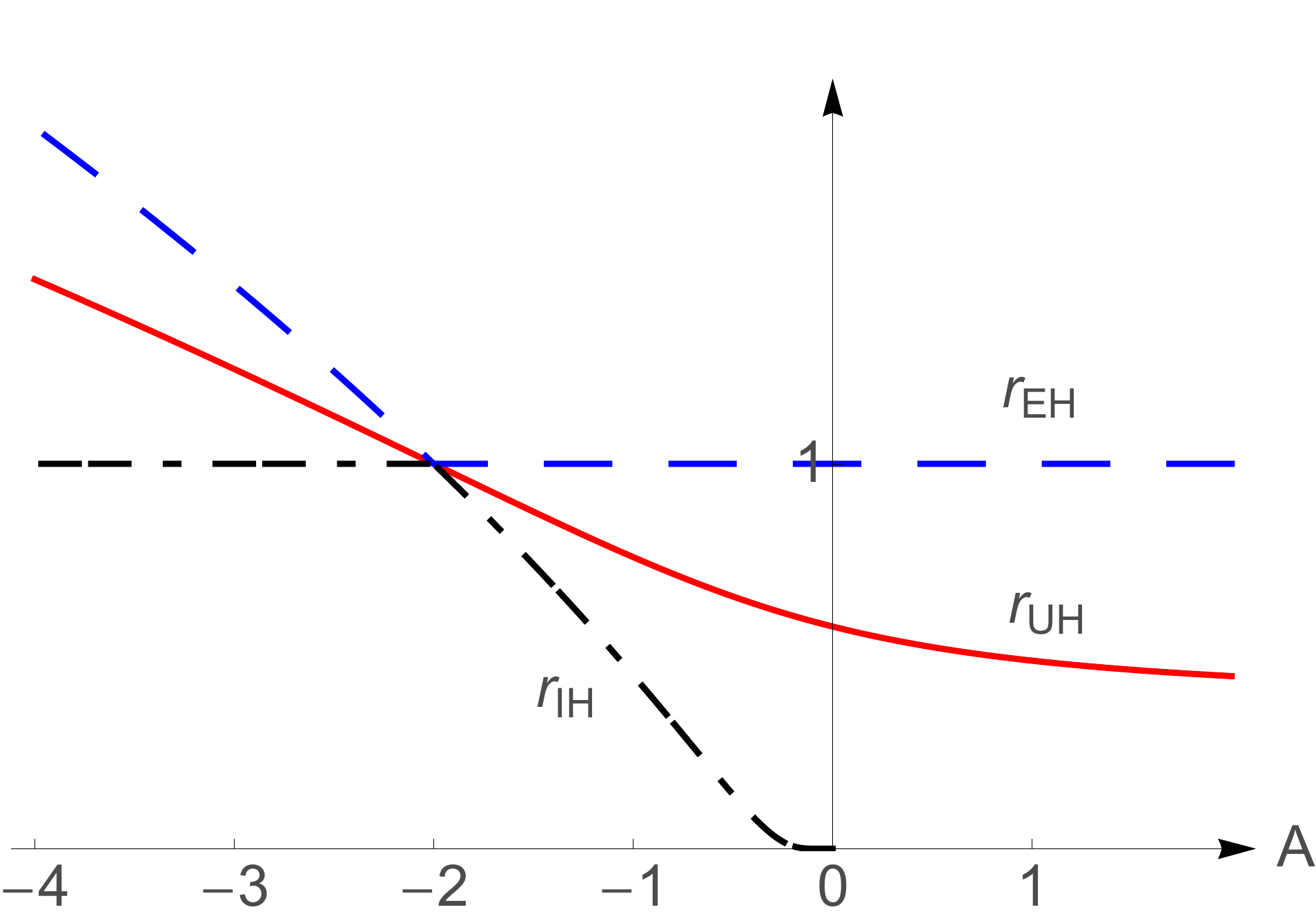}
\caption{The locations of the universal horizon $r = r_{UH}$, Killing (event) horizon $r = r_{EH}$, and  inner horizon $ r =
r_{IH}$  for  the solutions given by  Eqs.(\ref{3.23a}) and (\ref{3.23}) for $M = B > 0$.  When drawing this figure, we set $(M, B) = (1, 1)$.} \label{h3b}
\end{figure}

\begin{figure}[tbp]
\centering
\includegraphics[width=8cm]{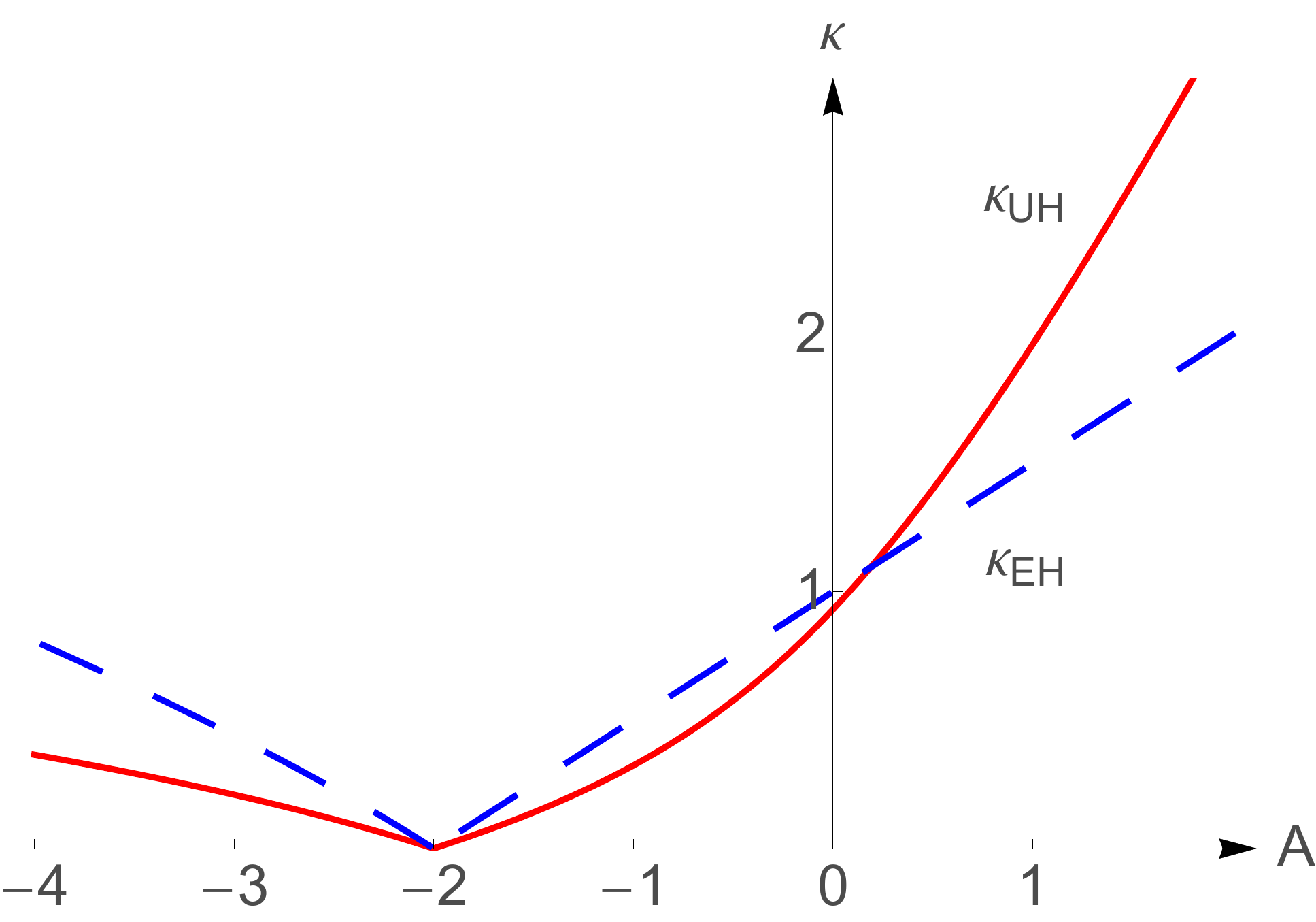}
\caption{The surface gravity $\kappa_{UH}$ on the universal horizon $r = r_{UH}$, and  the surface gravity $\kappa_{EH}$ on the event horizon
 $r = r_{EH}$  for the solutions given by Eqs.(\ref{3.23a}) and (\ref{3.23}) for $M =B > 0$. When drawing this figure, we set  $(M, B) = (1, 1)$.} \label{k3b}
\end{figure}

\begin{figure}[tbp]
\centering
\includegraphics[width=8cm]{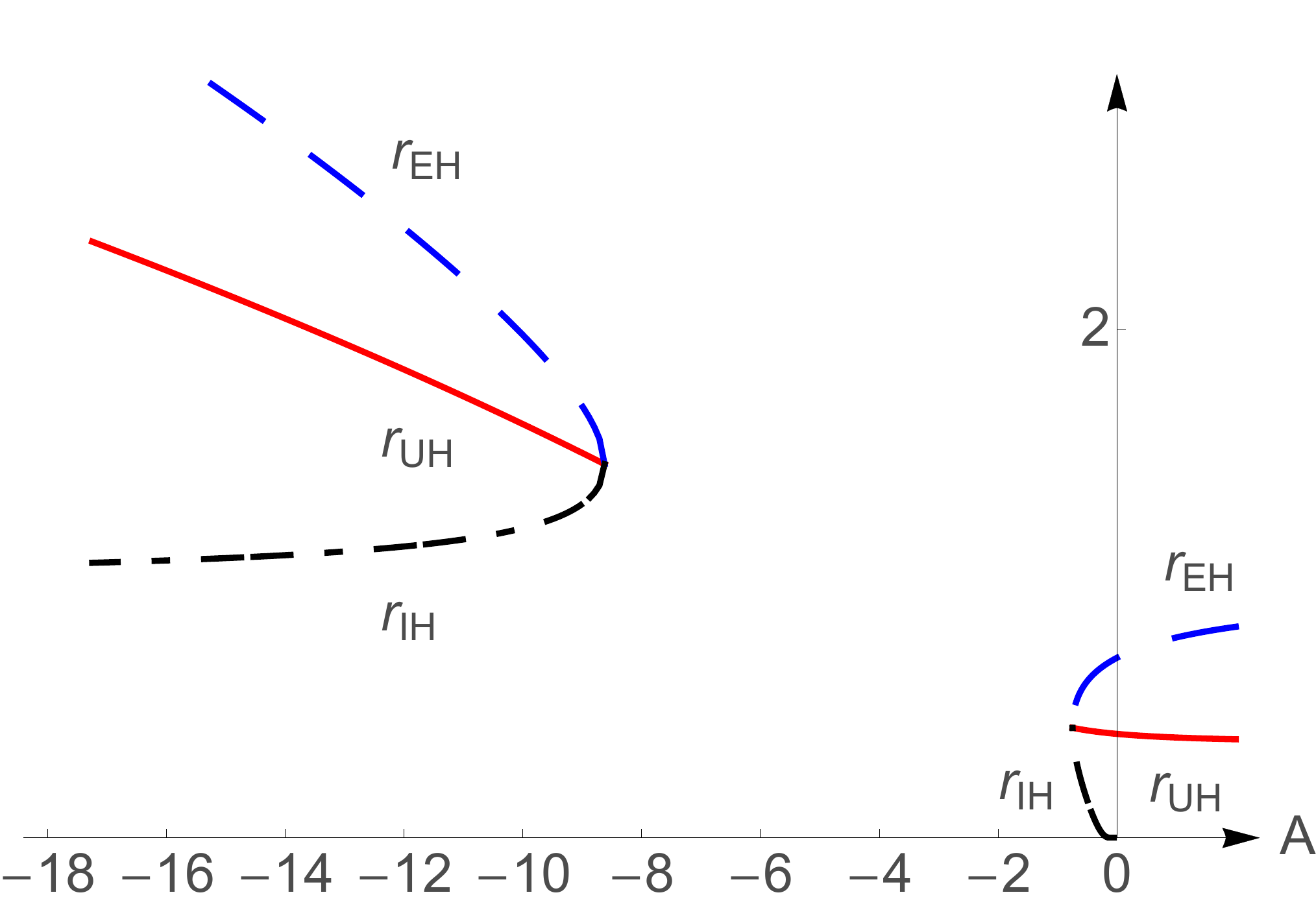}
\caption{The locations of the universal horizon $r = r_{UH}$, Killing (event) horizon $r = r_{EH}$, and  inner horizon $ r =
r_{IH}$  for  the solutions given by  Eqs.(\ref{3.23a}) and (\ref{3.23}) for $B > M > 0$.  When drawing this figure, we set $(M, B) = (1, 2)$.} \label{h3c}
\end{figure}

\begin{figure}[tbp]
\centering
\includegraphics[width=8cm]{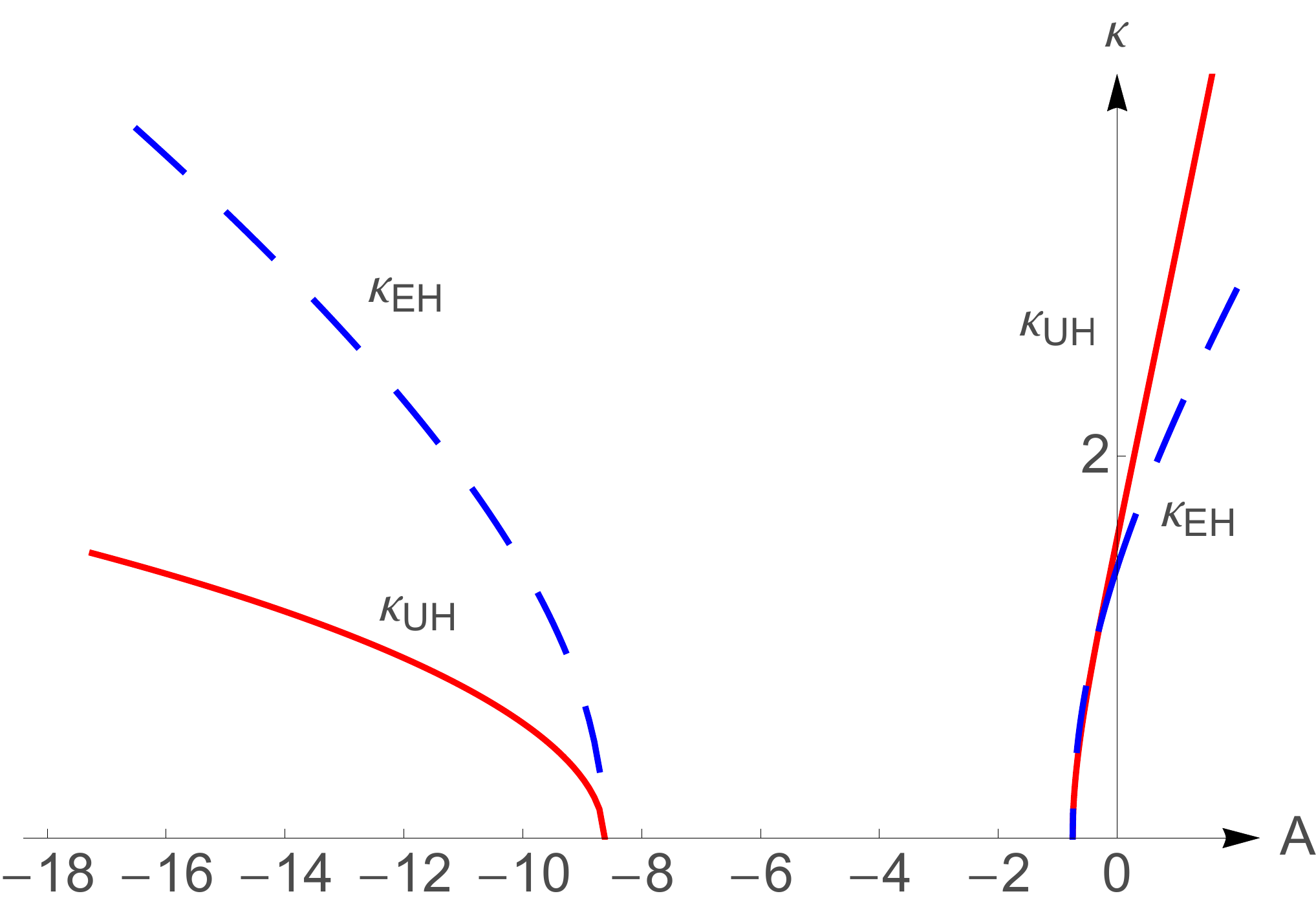}
\caption{The surface gravity $\kappa_{UH}$ on the universal horizon $r = r_{UH}$, and  the surface gravity $\kappa_{EH}$ on the event horizon
 $r = r_{EH}$  for the solutions given by Eqs.(\ref{3.23a}) and (\ref{3.23}) for $B> M > 0$. When drawing this figure, we set  $(M, B) = (1, 2)$.} \label{k3c}
\end{figure}

When $B > 0$, it is found that the locations of the universal horizons are different depending on whether  $M > B, M = B$  or  $M < B$.
 In particular, when $M > B$, Fig.\ref{h3a} shows the locations of the universal, Killing and inner horizons with different values of $A$, while
Fig.\ref{k3a} shows the surface gravity $\kappa_{UH}$ on the universal horizon and $\kappa_{EH}$ on
the event horizon  $r = r_{EH}$. On the other hand, Figs.\ref{h3b} and \ref{k3b} are for $ M =B$, and Figs.\ref{h3c} and \ref{k3c} for $M < B$.
Again, from these figures we can see that the  universal horizons are always inside the outer Killing horizon and outside the inner Killing horizon.
However, unlike the previous cases, now the surface gravity at $r =r_{UH}$ can be either larger or smaller than that   at $r = r_{EH}$.

It should be noted that for $B > 0$ there exist two Killing horizons only when $A \le 0$. Then,  similar to the case $d\ge 2$,
 the universal horizon coincides with the two Killing  horizons in the extremal  case, where the three horizons coincide. 
At this point, the parameter $A$ takes its extremal  value $A_c$,  given by, 
 \bqn
\lb{UHA5a} 
A_c \ln\left(\sqrt{\frac{-A_c}{2B}}\right) -\frac{A_c}{2}-M =0.
\eqn
In particular, when $M=B$, the above equation yields  $A_c=-2M$.

\begin{figure}[tbp]
\centering
\includegraphics[width=8cm]{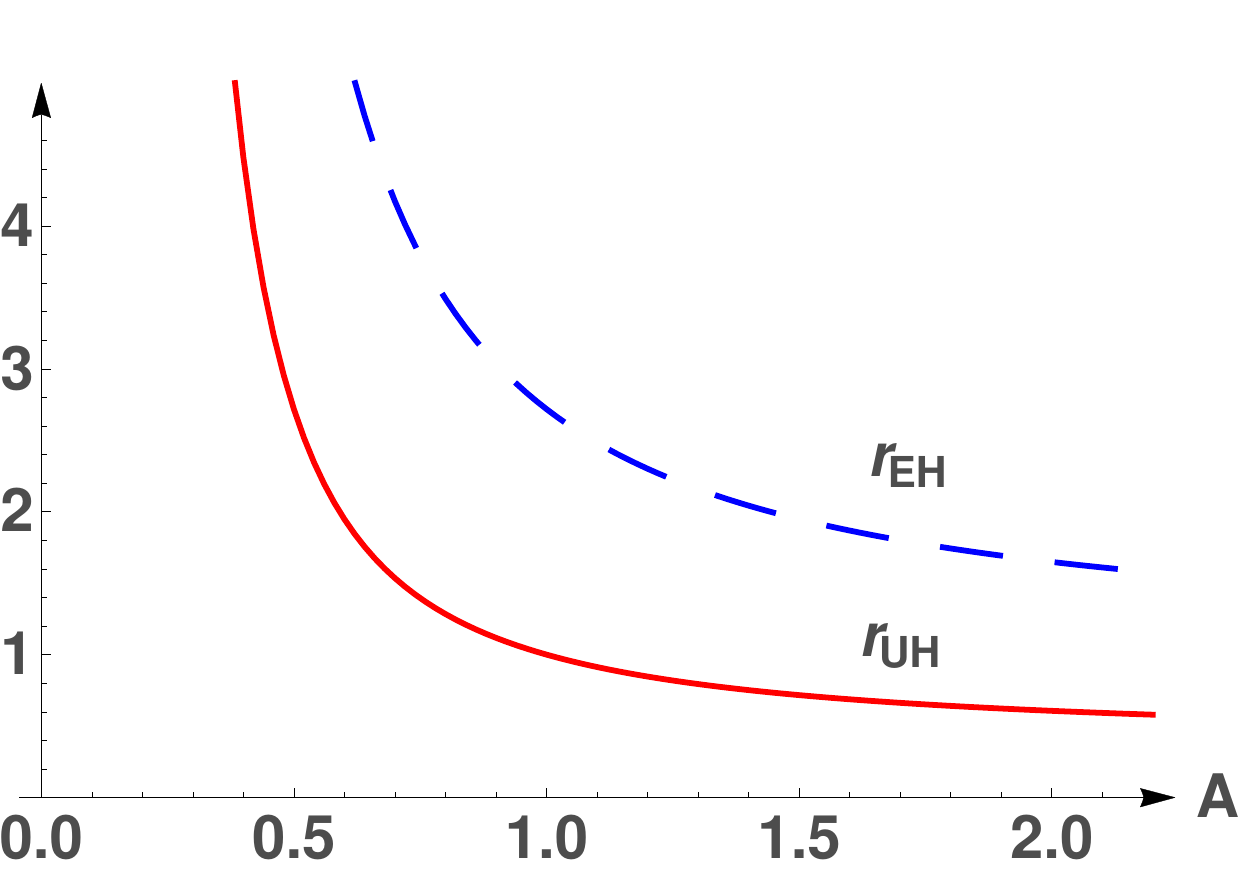}
\caption{The locations of the universal horizon $r = r_{UH}$ and Killing (event) horizon $r = r_{EH}$   for  the solutions given by  Eqs.(\ref{3.23a}) and
(\ref{3.23}) for $B = 0$.   When drawing this figure, we set $M=1$.} \label{hfb}
\end{figure}

\begin{figure}[tbp]
\centering
\includegraphics[width=8cm]{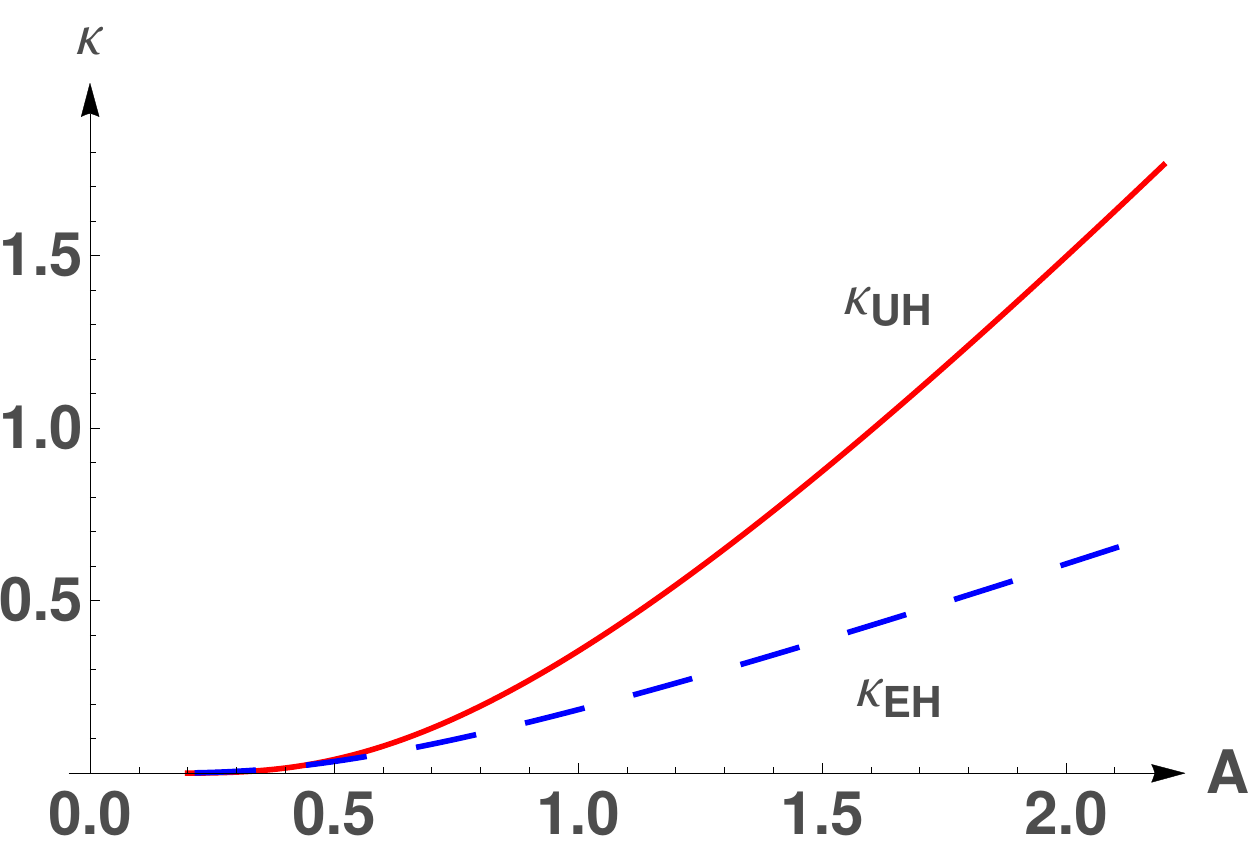}
\caption{The surface gravity $\kappa_{UH}$ on the
universal horizon $r = r_{UH}$, and  the surface gravity
$\kappa_{EH}$ on the event horizon
 $r = r_{EH}$  for the solutions given by Eqs.(\ref{3.23a}) and (\ref{3.23}) for $B = 0$. When drawing this figure, we set $M=1$.} \label{kfb}
\end{figure}

When $B = 0$, there exists only one Killing horizon. Fig.\ref{hfb} shows the locations of the universal and Killing  horizons  with different values of $A$,  while
Fig.\ref{kfb} shows the surface gravity $\kappa_{UH}$ on the universal horizon and $\kappa_{EH}$ on
the event horizon  $r = r_{EH}$. It is interesting to note that now the surface gravity on the universal
horizon again becomes  always larger than that on the outer Killing horizon.

When $B <0$,  the corresponding solutions are asymptotically de Sitter, and the Killing vector $\zeta^{\mu} = \delta^{\mu}_{t}$ becomes
 spacelike, $\zeta^{\lambda} \zeta_{\lambda}  = - F(r) \rightarrow \infty$, as $r \rightarrow \infty$. Then, it is not clear how to impose the boundary condition
for the khronon at  $ r  =  \infty$.

 \section{Conclusions}

In this paper, we have considered the definition of the universal horizons in any gravitational theories that violate the Lorentz symmetry, first discovered in the study
of the black hole thermodynamics in the khronometric theory \cite{BS11}. The latter is  equivalent to the Einstein-aether theory with the hypersurface-orthogonal condition
\cite{Jacobson10}. We have found that such a generalization is straightforward, and can be realized by simply promoting the khronon field to a probe field, and has no
effects on a given background, quite similar to a Killing field. This is in contrast to all the studies of the universal horizons carried out so far, as in all these studies the
khronon was always considered as part of the gravitational field \cite{UHsA,UHsB,BBM,CLMV,SVV}, although in some cases its backreactions    were assumed to be
negligible \cite{BS11}.

To apply the definition of the universal horizons  to theory with Lorentz violation, we have first studied static charged
spacwetimes in the non-projectable HL gravity in the IR limit in Sec. III, and found various  solutions. Some of them represent Lifshitz space-times with hyperscaling
violation, and some have structures with the presence of Killing horizons. In Sec. IV, we have solved the khronon equation in the
space-times that have Killing horizons, and found that universal horizons always exist inside these Killing horizons. Then, the regions inside these universal horizons naturally define
black holes in the HL gravity. We have also calculated  the surface gravity on  these universal horizons, by using the new definition given in \cite{CLMV}, which seems to be consistent with
the first law of black hole mechanics. We have shown explicitly that it can be either larger or smaller than the surface gravity on the Killing horizons, depending  on the particular
space-time backgrounds  considered. It would be very interesting to study the laws of black hole mechanics on the universal horizons found above.
Specially what is the role that the charge shall play, and how to define the entropy
of these black holes?

In addition, from the definition of the universal horizons one can see that the locations of them usually depend on three free parameters out of the four, $c_1, ..., c_4$. When
the khronon is part of the gravitational field, each of them  has its  physical interpretations \cite{EA}. What are their physical interpretations when the khronon is considered as a probe
field?  We wish to return to these important issues soon.

Finally, we would  like to note  that, although in our current paper
we have considered black holes only in the IR limit of the
non-projectable HL theory \cite{BPS}, such defined black holes also exist in
the full theory of the HL gravity. To see this, let us return to the
Schwarzschild black hole given by Eq.(\ref{eq1.1}). Blas and
Sibiryakov showed that  a universal horizon always  exist in such a
background  \cite{BS11}. Since the universal horizon is defined in
the covariant  form (\ref{eq1.2}),  if it exists in the   ($v,
r$)-coordinates, it must also exist in the  Painleve-Gullstrand
coordinates. The  Schwarzschild solution  given by Eq.(\ref{eq1.11})
in the Painleve-Gullstrand coordinates is a solution of the HL
gravity not only in the IR limit, but also in the full theory of the
HL gravity \cite{GLLSW}. This is simply  because the Ricci tensor
$R_{ij}$ of the surfaces $t = $ Constant in these   coordinates
vanishes identically, so all high-order operators made of
$R_{ij}$ have no contributions. Therefore, if it is a solution in
the IR limit, it must be also a solution of the full theory. Then,
the universal horizon found in the IR limit in \cite{BS11} is also the  universal
horizon of the full theory.

\section*{Acknowledgements}

This work is supported in part by DOE, DE-FG02-10ER41692 (AW);
Ci\^encia Sem Fronteiras, No. 004/2013-DRI/CAPES (AW);
NSFC No. 11375153 (AW); FAPESP No. 2012/08934-0 (EA, KL); CNPq (EA, KL);
and  NSFC No.10821504 (RC), No.11035008 (RG), and No.11375247 (RG).

\section*{Appendix A: Field Equations of the non-projectable HL Gravity in $(D+1)$-dimensions}

\renewcommand{\theequation}{A.\arabic{equation}} \setcounter{equation}{0}

In this Appendix, we shall first give a brief introduction to the non-projectable HL theory  in $(D+1)$-dimensions, and then consider the constraints from stability and ghost-free conditions.
To these goals,  let us start with the  ADM variables,
 \bq
\lb{2.0} \left(N, N_i, g_{ij}\right),\; (i,\; j = 1, 2,
\cdot\cdot\cdot, d+1),
 \eq
which are all functions of both $t$ and $x^i$, where $D = d+1$.
Then, the general action of the HL theory without the projectability
condition in $(D+1)$-dimensional spacetimes is given by
 \bqn
  \lb{2.1}
S &=& \zeta^2\int dt d^{D+1}x N \sqrt{g} \Big({\cal{L}}_{K} -
{\cal{L}}_{{V}}   +{\zeta^{-2}} {\cal{L}}_{M} \Big), ~~~~
 \eqn
where $ g={\rm det}(g_{ij})$, $\zeta^2 = {1}/{(16\pi G)}$, and
 \bqn
\lb{2.2a}
&& {\cal{L}}_{K} = K_{ij}K^{ij} -   \lambda K^{2},\nb\\
&& K_{ij} =  \frac{1}{2N}\left(- \dot{g}_{ij} + \nabla_{i}N_{j} +
\nabla_{j}N_{i}\right),
 \eqn
 where $\lambda$ is a dimensionless
coupling constant.  ${\cal{L}}_{{M}}$ is the Lagrangian of the electromagnetic field, which is given by \cite{KM,KKb}
 \bqn
\lb{actionMatter}
{\cal{L}}_{M}&=&\frac{\zeta^2}{4g_e^2}\Big[2\frac{g^{ij}}{N^2}\left({\cal F}_{0i}-N^k{\cal F}_{ki}\right)\left({\cal F}_{0j}-N^l{\cal F}_{lj}\right)\nb\\
&&-{\cal F}^{ij}{\cal F}_{ij}-{\cal G}\Big],
 \eqn
where $g_e$ is a coupling constant,  $B^i=\frac{1}{2}\frac{\epsilon^{ijk}}{\sqrt{g}}{\cal F}_{jk}$, $\epsilon^{ijk}$ is the Levi-Civita symbol,
${\cal F}_{ij}\equiv\nabla_i A_j-\nabla_j A_i$, ${\cal F}_{0i}\equiv\partial_t A_i-\partial_i A_0$ and the potential of the electromagnetic field is
  \bqn
\lb{potentialMatter}
{\cal G}=\alpha_0+\alpha_1 B_iB^i+\alpha_2 a_iB^i+ {{\cal{G}}}^{z>1},
 \eqn
where $\alpha_i$ is the arbitrary function of $A^iA_i$,  and ${{\cal{G}}}^{z>1}$ includes all of the operators  of ${\cal G}$ higher than order two.

 The potential $ {\cal{L}}_{V}$ can be cast in the form \cite{BPS,Horava,ZWWS},
 \bqn
  \lb{2.2}
 {\cal{L}}_{V} &=&  \gamma_{0}\zeta^{2}  + \beta  a_{i}a^{i}+ \gamma_1R
+ {{\cal{L}}}_{V}^{z>1},
 \eqn
where $\beta \equiv - \beta_0$, and ${{\cal{L}}}_{V}^{z>1}$ includes
all of  the higher-order derivative  terms of ${\cal{L}}_{V}$, and
 \bqn
 \lb{2.3}
a_i &=& \frac{N_{,i}}{N},\;\;\; a_{ij} = \nabla_{i}a_j.
 \eqn

 \subsection{Field Equations}

Variation of the action (\ref{2.1}) with respect to the lapse
function $N$ yields the Hamiltonian constraint
\bqn
\label{hami}
 {\cal{L}}_K + {\cal{L}}_V^R + F_V= 8\pi G J^t,\;\;
\eqn
where
\bqn
J^t&=& -\frac{\zeta^2}{8g_e^2}\Big[\frac{2g^{ij}}{N^2}\left({\cal F}_{0i}-N^k{\cal F}_{ki}\right)\left({\cal F}_{0j}-N^l{\cal F}_{lj}\right)+{\cal G}\nb\\
&&+{\cal F}^{ij}{\cal F}_{ij}-\alpha_2a_iB^i-\nabla_i\left(\alpha_2B^i\right)]+ {J}^{t}_{z> 1},\\
{\cal{L}}_V^R &=& \gamma_0 \zeta^2+\gamma_1R+{{\cal{L}}}_V^{R, z> 1},\\
F_V&=&-\beta\left(2a^i_i+a_ia^i\right)+{F}^{z>1}_V,
 \eqn
where ${{\cal{L}}}_V^{R, z > 1}$, ${F}_V^{z>1}$ and ${J}^{t}_{ z> 1}$ are all made of spatial derivative terms  higher than order  2.

Variation with respect to the shift vector $N_i$ yields the momentum constraint
 \bqn\lb{momen}
\nabla_j \pi^{ij}=8\pi G J^i,
 \eqn
where
 \bqn
\pi^{ij}&\equiv&  -K^{ij}+\lambda K g^{ij},\nb\\
 J^{i}&=& \frac{\zeta^2g^{lj}}{2g_e^2N}\left({\cal F}_{0l}-N^k{\cal F}_{kl}\right){\cal F}_{ij}+{J}^i_{z>1}.
 \eqn
Similarly,   ${J}^i_{z>1}$ consists of  spatial derivative terms  higher than order  2.

The dynamical equations are obtained by varying with respect to $g_{ij}$,
\bqn \label{dyn}
\frac{1}{\sqrt{g}N} \frac{\partial}{\partial t}\left(\sqrt{g} \pi^{ij}\right)+2(K^{ik}K^j_k-\lambda K K^{ij})\nb\\
-\frac12g_{ij}{\cal{L}}_K+\frac{1}{N}\nabla_k (\pi^{ik}N^j+\pi^{kj}N^i-\pi^{ij}N^k)\nb\\
-F^{ij}-F^{ij}_a=8\pi G \tau^{ij},\;\;\;\;\;\;
\eqn
where
 \bqn
\lb{tauij}
\tau_{ij}&=&-\frac{\zeta^2}{8g_e^2}\Big\{\frac{2}{N^2}\left({\cal F}_{0i}-N^k{\cal F}_{ki}\right)\left({\cal F}_{0j}-N^l{\cal F}_{lj}\right)\nb\\
&&-2g^{lk}{\cal F}_{il}{\cal F}_{jk}-{\cal G}_ij-\frac{g_{ij}}{2}\Big[\frac{2}{N^2}g^{qp}\left({\cal F}_{0q}-N^k{\cal F}_{kq}\right)\nb\\
&&\times\left({\cal F}_{0p}-N^l{\cal F}_{lp}\right)-{\cal F}^{lk}{\cal F}_{lk}-{\cal G}\Big]\Big\}+ \tau_{ij}^{z>1}, \nb\\
 {\cal G}_{ij}&=&(\alpha_0'+\alpha_1' B_iB^i+\alpha_2' a_iB^i)A_iA_j\nb\\
 &&+\alpha_1B_iB_j+\alpha_2a_{(i}B_{j)},\nb\\
 \alpha_i'&=&\frac{\partial \alpha_i}{\partial (A_iA^i)},\nb\\
F^{ij}&\equiv&\frac{1}{\sqrt{g}N}\frac{\delta (-\sqrt{g}N
{\cal{L}}_V^R)}{\delta g_{ij}}\nb\\
            &=& -\Lambda g^{ij}+\gamma_1\left(R^{ij}-\frac{1}{2}Rg^{ij}\right)\nb\\
            &&+\frac{\gamma_1}{N}\left(g^{ij}\nabla^2N-\nabla^i\nabla^jN\right)+{F}^{ij}_{z>1},\nb\\
F^{ij}_a&\equiv&\frac{1}{\sqrt{g}N}\frac{\delta (-\sqrt{g}N
{\cal{L}}_V^a)}{\delta g_{ij}}\nb\\
               &=& \beta \left(a^ia^j-\frac{1}{2}g^{ij}a^ka_k\right)+{F}^{ij}_{a, z> 1},
 \eqn
 with ${F}^{ij}_{z>1}$, ${F}^{ij}_{a, z> 1}$ and $\tau_{ij}^{z>1}$ all made of operators  higher  than order  two
 of $F^{ij}$, $F^{ij}_a$ and $\tau_{ij}$, respectively.

Finally, the Maxwell equations are
 \bqn
\label{MaxwellA}
&&\nabla^i\left[\frac{1}{N}\left({\cal F}_{0i}-N^k{\cal F}_{ki}\right)\right]+{\cal E}^{z>1}=0,\\
\label{MaxwellB}
&&\partial_t\left[\frac{g^{ij}}{N}\left({\cal F}_{0j}-N^k{\cal F}_{kj}\right)\right]\nb\\
&&-\nabla_k\left[\frac{g^{ij}N^k-g^{kj}N^i}{N}\left({\cal F}_{0j}-N^l{\cal F}_{lj}\right)\right]+\nabla_j\left(N{\cal F}^{ij}\right)\nb\\
&&-\frac{N}{2}A^i(\alpha_0'+\alpha_1' B_lB^l+\alpha_2' a_lB^l)\nb\\
&&-\frac{\epsilon^{kji}}{4\sqrt{g}}\nabla_j\left[N(2\alpha_1B_k+\alpha_2a_k)\right]+{\cal
E}^i_{z>1} =0.
 \eqn
Again,  ${\cal E}^{z>1}$ and ${\cal E}^i_{z>1}$ are made of operators  higher  than order two.

In the infrared (IR) limit, all the quantities made of operators higher than order two can be ignored. This is the case that will be assumed in the rest of the paper.

\subsection{Stability and Ghost-free Conditions}

When $\gamma_0 = 0$, the above HL theory admits the Minkowski
space-time
 \bq
 \lb{2.15}
 \left(\bar{N}, \bar{N}_i,
\bar{g}_{ij}\right) = \left(1, 0, \delta_{ij}\right),
 \eq
 as a solution of the theory. Then, its linear perturbations reveals that
the theory has two modes \cite{GHMT}, one represents the spin-2
massless gravitons with a dispersion relation\footnote{It should be noted that in the $d=1$ case, the spin-2 gravitons do not exist, so Eq.(\ref{2.16}) holds only for
$d \ge 2$.},
 \bq
 \lb{2.16}
\omega_{T}^2 = -\gamma_1 k^2,
 \eq
  and the other represents the
spin-0 massless gravitons with
 \bq
 \lb{2.17}
 \omega_{S}^2 =
-\frac{\gamma_1(\lambda -1)}{(d+1)\lambda -
1}\left[d\left(\frac{\gamma_1}{\beta} -1\right)+1\right] k^2. \eq
The stability conditions of these modes requires
\bq \lb{2.18}
\omega_{T}^2 > 0,\;\;\; \omega_{S}^2 > 0,
 \eq
 for any given $k$.

On the other hand, the kinetic term of the spin-0 gravitons is
proportional to $(\lambda -1)/[(d+1)\lambda - 1]$ \cite{GHMT}, so
the ghost-free condition requires
 \bq
 \lb{2.19}
 \frac{\lambda
-1}{(d+1)\lambda - 1} \ge 0.
 \eq
Thus, depending on the values of $d$ and $\lambda$, the stability and ghost-free conditions take different forms.

\subsubsection{$\lambda \not=1,\; d \ge 2$}

 In this case, the conditions (\ref{2.18}) and (\ref{2.19}) require
\bqn
\lb{2.20a}
&&  \gamma_1 < 0, \;\;\;
 \frac{d\gamma_1}{d-1} < \beta < 0,\\
 \lb{2.20b}
 && i)\; \lambda >  1,\;\;\; {\mbox{or}} \;\;\;\; ii)\;
\lambda \le \frac{1}{d+1}.
 \eqn

\subsubsection{$\lambda =1,\; d \ge 2$}

When $\lambda =1$ and $d \ge 2$, we have $\omega_{S}^2 = 0$, and  the conditions (\ref{2.18}) and (\ref{2.19}) require 
\bq
\lb{2.20c} \gamma_1 < 0, \;\;\; {\mbox{and}} \;\; \beta \;\;  {\mbox{is free}}.
\eq
  In particular, $\beta$ can be zero, as   can be seen  from Eq.(\ref{2.17}).

\subsubsection{$\lambda \not=1,\; d =1$}

In this case,  the spin-2 gravitons do not exist, as noted above, and then we find that
\bq
\lb{2.20d}
\omega_S^2 = - \frac{\gamma_1^2(\lambda - 1)}{\beta(2\lambda -1)}k^2.
\eq
Thus, the stability and ghost-free conditions require
\bqn
\lb{2.20e}
&&
  \beta <  0,\;\;\;  \gamma_1\;\;  {\mbox{is free}},  \\
 \lb{2.20f}
 && i)\; \lambda >  1,\;\;\; {\mbox{or}} \;\;\;\; ii)\;
\lambda \le \frac{1}{2}.
 \eqn

\subsubsection{$\lambda =1 = d$}

In this case, from the above we can see that  both $\beta$ and $\gamma_1$ are free,
\bqn
\lb{2.20g}
&&
  \beta\;\; {\mbox{and }}  \;\;  \gamma_1\;\;  {\mbox{ are all free}}.
 \eqn

\end{document}